\documentclass[a4paper,11pt]{article}

\usepackage[english]{babel}
\usepackage[T1]{fontenc}
\usepackage[utf8]{inputenc}
\usepackage{libertine}
\usepackage{hyperref}
\hypersetup{
    colorlinks,
    linkcolor={red!50!black},
    citecolor={blue!50!black},
    urlcolor={blue!80!black}
}
\usepackage{graphicx}
\usepackage{systeme}
\usepackage{amsmath}
\usepackage{lastpage}
\usepackage{changepage}
\usepackage{geometry}
\usepackage{wasysym} 
\usepackage{tikz} 
\usepackage{schemabloc} 
\usepackage{amssymb}
\usepackage{gensymb}
\usepackage{amsfonts}
\usepackage{multicol}
\usepackage{listings}
\usepackage{etoolbox}
\usepackage[justification=centering]{caption}
\usepackage{xcolor}
\usepackage{textcomp}
\usepackage{caption}
\usepackage{fancyhdr}
\newcommand{\HRule}{\rule{\linewidth}{0.5mm}}

\newcommand{\mathd}{\mathrm{d}}

\newcommand{\uvec}[1]{\underline{#1}}
\newcommand{\umat}[1]{\underline{\underline{#1}}}

\begin{document}

\begin{center}
	
	{ \huge \bfseries Complex Orthogonal Decomposition (C.O.D.) \\using Python}\\[0.3cm]
	{\Large \textcolor{gray}{}}\\[0.1cm] 

    \large 
    Marc Vacher$^{*,1,2}$, Stéphane Perrard$^{2}$ and Sophie Ramananarivo$^{1}$\\[1em]
    \normalsize
    $^{1}$Laboratoire d'Hydrodynamique de l'Ecole polytechnique (LadHyX),\\
    Ecole polytechnique, Institut Polytechnique de Paris\\[0.5em]
    $^{2}$Laboratoire Physique et Mécanique des Milieux Hétérogènes (PMMH),\\
    ESPCI, Paris Sciences et Lettres\\[0.5em]
    $^{*}$E-mail: \textit{marc.vacher@polytechnique.edu}
	
	\HRule \\[0.5cm]
\end{center}

\section*{Abstract}

This work presents the application of the \textbf{Complex Orthogonal Decomposition (C.O.D.)} to a simple spatio-temporal signal. C.O.D. has been introduced first in the article of B. Feeny, entitled "A Complex Orthogonal Decomposition for Wave Motion Analysis" \cite{feeny_complex_2008}, published in the \textit{Journal of Sound and Vibration}.

The purpose of this signal analysis method is to extract spatial and temporal modes out of a signal. This approach is especially suited to deal with oscillatory signals where phase information is important and where spatial forms are unknown. 

We provide two theoretical chapters presenting the main mathematical concepts behind C.O.D. and a series of example (with associated  \textit{Python} scripts) to demonstrate the efficiency of the method and some characteristical features.

\setcounter{tocdepth}{1}
\tableofcontents

\newpage

\section{Motivation and presentation}

The main idea of Complex Orthogonal Decomposition (C.O.D.) is to analyse a signal of time and space $s(t,x)$ representing a physical quantity that oscillates. To introduce it, we look at a real example that demonstrates its purpose and how it works : fish swimming.

\begin{figure}[ht!]
    \centering
    \includegraphics[scale=0.78]{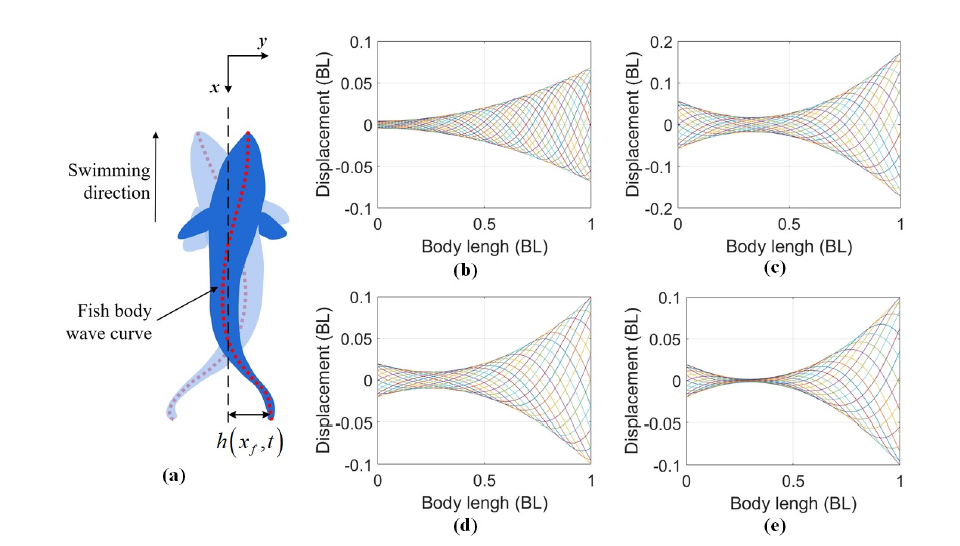}
    \caption[]{Figure from Chen \textit{et al.} \cite{chen_complex_2024} -  Fish body wave curve: (a) Diagram of the fish body wave curve. The blue area is the outline
of a BCF fish, and the red line segment is the fish body wave curve. (b)-(e) Fish body wave curves of four
different species of fish. Curves of different colors represent the body midline displacement of fish at 20 equally spaced time intervals during a single tail beat cycle.}
    \label{fig1}
\end{figure}

Indeed, many fishes have oscillatory motions when they swim, with definite frequencies linked to their species, their size or if they swim in schools or not for example \cite{chen_complex_2024}. However, this wave motion is not trivial : if we analyse their motion using the centreline of their body, researchers have shown that this oscillatory motion is not \textit{spatially} periodical at all. It means that the spatial form of the wave is not trivial, as we can see on Fig. \ref{fig1}. (b)-(e).

Mathematically, it means that the signal $s(t,x)$ (being the amplitude of oscillation of the fish, $t$ being time and $x$ a position along its centerline) does not decompose well on a 2D space-time Fourier basis. Consequently, the following decomposition 

\begin{equation}
    s(t,x) = \sum_{j=1}^{+\infty}A_j \cos(k_jx-\omega_jt+\phi_j)
\end{equation}

cannot be truncated at relatively low index. The space Fourier basis do not represent well the motion in this case. 

In this context, C.O.D. is the appropriate tool to analyse the signal. In a sense, it allows you to have your signal decomposed as :

\begin{equation}
s(t,x) = \sum_{k=1}^{+\infty} T_k(t) \, X_k(x),
\end{equation}

with $X_k$ being spatially orthogonal to each other. In the case of the fish, it is interesting to see if some frequencies are associated with specific unknown body motions that would enhance the efficiency of the swim. Additionally, the C.O.D. provides a tool called the travelling index, a number that quantifies how much a given mode resembles a travelling wave versus a standing wave. It is typically an indicator that can give clues on the swim efficiency \cite{anastasiadis_identification_2023}. 

From a mathematical point of view, the C.O.D. is not that different from the classical Proper Orthogonal Decomposition (P.O.D.) in terms of conceptual ideas \cite{feeny_complex_2008} but is not used as much in studies. This paper aims at recalling the theory concerning Complex Orthogonal Decomposition, with applicative examples, and demonstrate its efficiency. 

\textit{Python} scripts are attached to support these examples and give easy access to the method for research or pedagogical purposes. Are attached to this document:

\begin{itemize}
    \item a \textit{Python} package \textit{pack\_COD}, including all the related functions performing the Complex Orthogonal Decomposition, plus other verification and plotting tools.
    \item \textit{test\_COD\_ex\_1.py} : a first example to illustrate the main principles of the C.O.D. and how the package works. The theoretical analysis that goes along this case is developed in section \ref{IV}.
    \item \textit{test\_COD\_ex\_2.py} : this example shows that in the case of non perfectly periodical signals, some approximation in the method can lead to errors. We explain quantitively their origin in   section \ref{VI}
    \item \textit{test\_COD\_ex\_3.py} : this example shows the efficiency of C.O.D. if one spatial form is associated with a multi-frequency oscillations. We use a frequency modulated signal to demonstrate this fact. It corresponds to section \ref{VII}
\end{itemize}

In addition to each of these four examples, videos of the 2D spatio-temporal signals are associated. Three other examples are added :

\begin{itemize}
    \item \textit{test\_COD\_ex\_1\_nu.py} : this example shows how the C.O.D. works in the context of non-uniform space grids. It corresponds to the Appendice.
    \item \textit{test\_COD\_ex\_your\_data.py} : this script allows you to try the C.O.D. on your own data, put into the form of two \textit{.csv} files.
    \item \textit{test\_COD\_ex\_noise.py} : it is an example of how C.O.D. deals with noise. This example is not discussed since it is not the main purpose of the document here, yet interesting things could be analysed.
\end{itemize}

\newpage
\section{Spatio-temporal signal processing and C.O.D.}

Let \( s(t, x) \in \mathbb{R} \) be a real-valued spatio-temporal signal, smooth in both time \( t\in \Omega_t \) and space \( x \in \Omega_x \). $\Omega_t = [0,T]$, with $T\in \mathbb{\overline{R^{+}}}$ and $\Omega_x =
[x_{lim_1},x_{lim_2}]$, with $(x_{lim_1},x_{lim_2})$ in $\overline{\mathbb{R}}^2$. The core idea of C.O.D. is to express the real analytic signal as the real part of sum of separable complex spatio-temporal modes:

\begin{equation}
s(t, x) = \Re\left(\sum_{j=1}^{+\infty} a_j(t) \, \phi_j(x)\right)
\end{equation}

where:

\begin{itemize}
  \item \( \phi_j(x) \in \mathbb{C} \) are orthonormal spatial modes,
  \item \( a_j(t) \in \mathbb{C} \) are complex temporal coefficients (also called complex ortogonal coordinates, C.O.C. \cite{feeny_complex_2008}).
\end{itemize}

To do that, we need to have an equivalent complex signal. We use the Hilbert transform to achieve this goal.

\subsection*{1. Complex analytic signal and Hilbert transform}

Given a real-valued signal \( s(t,x) \), we define its \textbf{Hilbert transform} in time at each spatial position \( x \). This transform captures the oscillatory structure of the signal in a way that separates amplitude and phase.

Formally, the Hilbert transform \( \mathcal{H} \{ s(t,x) \} \) is defined by convolution in time \cite{king_hilbert_2008}:

\begin{equation}
\widetilde{s}(t,x) = \mathcal{H}\{s\}(t,x) = \frac{1}{\pi} \, \mathrm{p.v.} \int_{\Omega_t} \frac{s(\tau,x)}{t - \tau} \, \mathd \tau,
\end{equation}
where “p.v.” denotes the Cauchy principal value.

The complex analytic signal is then:
\begin{equation}
s_c(t,x) = s(t,x) + i \, \widetilde{s}(t,x).
\end{equation}

\textbf{Relation to the Fourier transform :}

In the frequency domain, constructing the complex signal corresponds to suppressing the negative frequency components and doubling the positive ones. This is efficiently done using the (complex) Fourier transform.

Let \( \mathcal{F}_t\{ s \}(f,x) = \hat{s}(f,x) \) be the temporal Fourier transform of \( s(t,x) \). Then the complex part of the signal can be obtained via:
\begin{equation}
\hat{s_c}(f,x) =
\begin{cases}
2\hat{s}(f,x), & f > 0, \\
\hat{s}(0,x), & f = 0, \\
0, & f < 0.
\end{cases}
\end{equation}

The complex analytic signal \( s_c(t,x) \) is then recovered by taking the inverse Fourier transform of \( \hat{s_c}(f,x) \):
\begin{equation}
s_c(t,x) = \mathcal{F}_t^{-1} \{ \hat{s_c}(f,x) \}.
\end{equation}

This frequency-domain filtering approach is equivalent to performing the Hilbert transform in the time domain, and is especially useful in numerical computations. It ensures that the resulting \( s_c(t,x) \) contains only positive frequencies, making it ideal for phase and envelope analysis.

\subsection*{2. Modal decomposition via C.O.D.}

The spatial modes \( \phi_j(x) \) are obtained by solving the eigenvalue problem of the spatial covariance operator:

\begin{equation}
\mathcal{R}[f](x) = \frac{1}{T} \int_0^T \left( \int_{x_{lim_1}}^{x_{lim_2}} s_c(t, x') f(x')^* \mathd x' \right) s_c(t, x) \, \mathd t,
\end{equation}

where \( ^* \) denotes complex conjugation.

This operator is Hermitian and positive semi-definite, ensuring a complete orthonormal set of eigenfunctions.

We call :\begin{itemize}
    \item $\lambda_j^e$ the eigenvalues (that are real and positive, because of the positive semi-definite characteristics of the operator)\footnote{We use the exponent $^{e}$ to avoid any confusion with $\lambda$ - the spatial wavelength.}.
    \item $\phi_j$ the associated eigenvectors.
\end{itemize}

The temporal coefficients \( a_j(t) \) can be retrieved by projection:
\begin{equation}
a_j(t) = \int_{\Omega_x} s_c(t,x) \, \phi_j(x)^* \, dx
\end{equation}

These properties follow from the fact that the spatial modes are eigenfunctions of a Hermitian (self-adjoint) covariance operator.

\subsection*{3. Reconstruction of the real signal}

Once the complex signal \( s_c(t,x) \) has been decomposed as
\begin{equation}
s_c(t,x) = \sum_{j=1}^{+\infty} a_j(t) \, \phi_j(x),
\end{equation}
where \( a_j(t) \) are complex-valued temporal coefficients and \( \phi_j(x) \) are complex-valued spatial modes (orthonormal in \( x \)), the original real-valued signal \( s(t,x) \) is reconstructed by taking the real part:
\begin{equation}
s(t,x) = \Re\left[ s_c(t,x) \right] = \Re\left[ \sum_{j=1}^{+\infty} a_j(t) \, \phi_j(x) \right].
\end{equation}

Expanding the real part explicitly, using \( a_j(t) = a_j^{\Re}(t) + i a_j^{\Im}(t) \) and \( \phi_j(x) = \phi_j^{\Re}(x) + i \phi_j^{\Im}(x) \), we get:
\begin{equation}
s(t,x) = \sum_{j=1}^{+\infty} \left[ a_j^{\Re}(t) \, \phi_j^{\Re}(x) - a_j^{\Im}(t) \, \phi_j^{\Im}(x) \right].
\end{equation}

\textbf{Expressing the sum as products of real functions:}

In general, it is \emph{not} possible to express each term \( \Re[a_j(t) \phi_j(x)] \) as a single product of one function of time and one function of space unless either \( a_j(t) \) or \( \phi_j(x) \) is real.

However, we can rewrite the signal as a sum of products of real functions by defining:
\begin{equation}
T_{j}^{(1)}(t) := a_j^{\Re}(t), \quad X_{j}^{(1)}(x) := \phi_j^{\Re}(x),
\end{equation}
\begin{equation}
T_{j}^{(2)}(t) := -a_j^{\Im}(t), \quad X_{j}^{(2)}(x) := \phi_j^{\Im}(x).
\end{equation}

Then the real signal becomes:
\begin{equation}
s(t,x) = \sum_{n=1}^{+\infty} \left[ T_{j}^{(1)}(t) \, X_{j}^{(1)}(x) + T_{j}^{(2)}(t) \, X_{j}^{(2)}(x) \right].
\end{equation}

By reindexing, e.g., setting \( k = 2j - 1 \) and \( k = 2j \), one can write:
\begin{equation}
s(t,x) = \sum_{k=1}^{+\infty} T_k(t) \, X_k(x),
\end{equation}
with all \( T_k(t) \) and \( X_k(x) \) real-valued functions. This expresses the signal as a sum of separable real modes.

\subsection*{4. Modal energies}

The energy associated with each mode is given by the eigenvalues:

\begin{equation}
\lambda_j^e = \frac{1}{T} \int_0^T |a_j(t)|^2 \, \mathd t.
\end{equation}

Modes are sorted so that \( \lambda_1^e \geq \lambda_2^e \geq \cdots \), with the first few often capturing most of the signal energy. Using Parseval identity, we also have the link with the Power Spectral Density : 

\begin{equation}
     \frac{1}{T} \int_0^T |a_j(t)|^2 \, \mathd t = \frac{2\pi}{T} \int_{f_{min}}^{f_{max}} |\hat{a_j}|(\omega)^2\mathd \omega
\end{equation}

\subsection*{5. Orthogonality and norm of the spatial modes}

The spatial modes \( \phi_j(x) \) obtained through Complex Orthogonal Decomposition (C.O.D.) form an \textbf{orthonormal basis} in the space of square-integrable complex functions over the spatial domain.

\paragraph*{Orthogonality:}
For all \( j \neq k \), the modes are orthogonal:
\begin{equation}
\langle \phi_j, \phi_k \rangle = \int_{\Omega_x} \phi_j(x)^* \phi_k(x) \, dx = 0
\end{equation}

with $^*$ denoting the complex conjugate of the quantity.

\paragraph*{Normalization:}
Each mode is normalized to unit norm:
\begin{equation}
\| \phi_j \|^2 = \int |\phi_j(x)|^2 \, dx = 1
\end{equation}

This orthonormality ensures that the modal decomposition
\begin{equation}
s_c(t, x) = \sum_{j=1}^{+\infty} a_j(t) \, \phi_j(x)
\end{equation}
is \textbf{unique}.

\newpage

\subsection*{5. Travelling index}

Each spatial mode \( \mathbf{\phi}_j(x) \) is a complex-valued function defined over the spatial domain \( \Omega_x \). The goal of the travelling index is to quantify how much a given mode resembles a travelling wave versus a standing wave. This is achieved by defining a \textbf{travelling index} \( \alpha_j \in [0,1] \).

Let the \( j \)-th mode be written in terms of its real and imaginary components:

\begin{equation}
\mathbf{\phi}_j(x) = \Re[\mathbf{\phi}_j(x)] + i \Im[\mathbf{\phi}_j(x)] = \phi_j^{\Re}(x) + i \phi_j^{\Im}(x),
\end{equation}

We compute the complex matrix $W_j$ and its Gram matrix $G_j$ :

\begin{equation}
W_j(x) = 
\begin{bmatrix}
\phi_j^{\Re}(x) & \phi_j^{\Im}(x)
\end{bmatrix}
\in \mathbb{R}^{1\times2}
\end{equation}

\begin{equation}G_j(x) = \begin{bmatrix}
\langle \phi_j^{\Re}(x), \phi_j^{\Re}(x) \rangle & \langle \phi_j^{\Re}(x), \phi_j^{\Im}(x) \rangle \\
\langle \phi_j^{\Im}(x), \phi_j^{\Re}(x) \rangle & \langle \phi_j^{\Im}(x), \phi_j^{\Im}(x) \rangle
\end{bmatrix} =
\begin{bmatrix}
a & c \\
c & b
\end{bmatrix}
\end{equation}

Then we define:
\begin{equation}
\alpha_j = \frac{1}{\kappa(W_j)} = \frac{\sigma_{\min}}{\sigma_{\max}} = \sqrt{\frac{\lambda_{min}^{G}}{\lambda_{max}^{G}}}
\end{equation}
where \( \kappa(W) \) is the condition number of the matrix \( W \), and \( \sigma_{\min} \), \( \sigma_{\max} \) are the smallest and largest singular values. In this context, the singular values of $W_j$ are the square roots of $G_j$ eigenvalues $\lambda^{G}$ \cite{golub_vanloan_2013}.

We use the quadratic formulas for the eigenvalues being the roots of the characteristic polynomial of the matrix since $G_j$ matrixes has a 2 by 2 size \cite{meyer_2000}. The travelling index \( \alpha_j \) is then computed as :

\begin{equation}
\alpha_j  = 
\sqrt{\frac{a + b - \sqrt{(a - b)^2 + 4c^2}}{a + b + \sqrt{(a - b)^2 + 4c^2}}
}
\end{equation}

\begin{itemize}
    \item \( \alpha_j = 1 \): the mode corresponds to a perfectly travelling wave (circular trace in the complex plane).
    \item \( \alpha_j = 0 \): the mode is purely standing (real or imaginary only).
\end{itemize}

The travelling index thus provides a physically interpretable scalar measure of wave behavior within each mode extracted by C.O.D.

\newpage

\section{Discrete equivalent for C.O.D.}

In reality, we often work with discrete signals coming from experiment. We consider a discretely sampled spatio-temporal signal represented by the matrix \( \umat{S} \in \mathbb{R}^{N_t \times N_x} \), where:
\begin{itemize}
    \item \( N_t \) is the number of time steps,
\item \( N_x \) is the number of spatial points,
\item each row of \( \umat{S} \) corresponds to a fixed time \( t_n \),
\item each column corresponds to a fixed spatial location \( x_j \).
\end{itemize}
In the following, $j \in [1,N_x]$ and $n \in [1,N_t]$.

\subsection*{1. Complex signal matrix via Hilbert transform}

To form the analytic complex signal, we apply a discrete Hilbert transform in the temporal direction (row-wise). For each spatial point \( x_j \), we define the complex signal :

\begin{equation}
(\uvec{s_c}){_j} = \uvec{s}_j + i\, \mathcal{H}_d[\uvec{s}_j],
\end{equation}

where: \begin{itemize}
    \item \( \uvec{s}_j \in \mathbb{R}^{N_t\times1} \) is the time signal at location \( x_j \) (i.e., column \( j \) of \( \umat{S} \)),
    \item \( \mathcal{H}_d \) denotes the discrete Hilbert transform.
\end{itemize}

This yields the complex signal matrix \( \umat{S_c} \in \mathbb{C}^{N_t \times N_x} \):

\begin{equation}
\umat{S_c} = \umat{S} + i\, \widetilde{\umat{S}},
\end{equation}

where \( \widetilde{\umat{S}} \) contains the Hilbert transforms of the columns of \( \umat{S} \). To get it, we use the same relation with the Fourier transform described in the continuous case, instead here one should use the Fast Fourier Transform to have discrete equivalent.

\textbf{Relation to the FFT :}

Let \( \text{FFT}\{ \umat{S} \} = \widehat{\umat{S}} \) be the discrete temporal Fourier transform of the matrix $\umat{S}$\footnote{As written here, the complex FFT has to be shifted so that the frequency domain between $[-\frac{N_t}{2T},\frac{N_t}{2T}]$ is mapped in order by $n \in [0,N_t]$}. Then the complex part of the signal can be obtained via the following matrix:
\begin{equation}
(\widehat{\umat{S_c}})_{nj} =
\begin{cases}
2\widehat{\umat{S}}_{nj}, & n > \lfloor N_t/2 \rfloor, \\
\widehat{\umat{S}}_{nj}, & n = \lfloor N_t/2\rfloor, \\
0, & n <  \lfloor N_t/2\rfloor .
\end{cases}
\end{equation}

The complex matrix \( \umat{S_c} \) is then recovered by taking the complex inverse Fourier transform of \( \widehat{\umat{S_c}} \):
\begin{equation}
\umat{S_c} = \text{FFT}^{-1} \{ \widehat{\umat{S_c}} \}.
\end{equation}

\newpage

\subsection*{2. Modal decomposition via C.O.D.}

We now transpose \( \umat{S_c} \), resulting in:

\begin{equation}
\umat{Z} = \umat{S_c}^T \in \mathbb{C}^{N_x \times N_t},
\end{equation}

We define the temporal covariance matrix:

\begin{equation}
\umat{R} = \frac{1}{N_t} \umat{Z} \umat{Z}^\dagger \in \mathbb{C}^{N_x \times N_x},
\end{equation}

which is Hermitian and positive semi-definite. $\umat{Z}^\dagger$ is the hermitian transpose of $\umat{Z}$, so that $\umat{Z}^\dagger = (\umat{Z}^{\ast})^T$

We solve the eigenvalue problem:

\begin{equation}
\umat{R} \uvec{\phi}_j = \lambda_j^e \uvec{\phi}_j,
\end{equation}

yielding:
- eigenvalues \( \lambda_j^e \in \mathbb{R}_+ \) representing the modal energies,
- eigenvectors \( \uvec{\phi}_j \in \mathbb{C}^{N_x\times 1} \) representing the spatial modes.

The temporal coefficients \( \uvec{a}_j \in \mathbb{C}^{1 \times N_t} \) are obtained via projection:

\begin{equation}
\uvec{a}_j = \uvec{\phi}_j^\dagger \umat{Z}
\end{equation}

or, equivalently, stacking the projections gives:

\begin{equation}
\umat{A} = \umat{\Phi}^\dagger \umat{Z},
\end{equation}

where \( \umat{\Phi} = [\uvec{\phi}_1, \uvec{\phi}_2, \dots] \in \mathbb{C}^{N_x \times N_x} \) is the matrix of spatial modes and \( \umat{A} \in \mathbb{C}^{N_x \times N_t} \) contains the modal time series.

Thus, the dataset can be reconstructed as a sum of contributions:
\begin{equation}
\umat{S_c} = \umat{A}^{T}\umat{\Phi}^{T} = \sum_{j=1}^{N_x} \uvec{a_j}^{T}\uvec{\phi_j}^{T}
\end{equation}

This decomposition is energy-preserving due to the orthonormality of the modes.

\subsection*{3. Reconstruction of the real matrix}

To recover the real-valued signal \( \umat{S} \), take the real part of the transpose of \( \umat{S_c} \):

\begin{equation}
\umat{S} \approx \Re[ \umat{Z}^T ] = \Re[ \umat{S_c} ].
\end{equation}

\subsection*{4. Modal energies}

Each eigenvalue \( \lambda_j \) quantifies the energy of mode \( j \):

\begin{equation}
\lambda_j^e = \frac{1}{N_t} \| \uvec{a}_j \|^2 
\end{equation}

The norm \( \| \cdot \| \) refers to the standard Euclidean norm (or \( L^2 \)-norm) applied to complex vectors, defined by:

\begin{equation}
\| \uvec{a}_j \|^2 = \sum_{n=1}^{N_t} |a_j(t_n)|^2 = \sum_{n=1}^{N_t} |a_{jn}|^2 ,
\end{equation}

Modes are typically sorted by decreasing \( \lambda^{e}_j \) to identify the most energetic structures.

\subsection*{5. Orthogonality and norm of the spatial modes}

We have the same discrete equivalent properties as for the continuous C.O.D.

\paragraph*{Orthogonality:}
The spatial modes \( \uvec{\phi}_j \) form an orthonormal set in \( \mathbb{C}^{N_x\times1} \). For \( j \neq k \), we have:
\begin{equation}
\uvec{\phi}_j^\dagger \uvec{\phi}_k = 0
\end{equation}

\paragraph*{Normalization:}
Each mode is normalized to unit norm:
\begin{equation}
\| \uvec{\phi}_j \|^2 = \uvec{\phi}_j^\dagger \uvec{\phi}_j = 1
\end{equation}

These properties arise from the Hermitian nature of the matrix \( \umat{R} \), and they ensure that the modes form a complete orthonormal basis for the spatial structure of the data.

\subsection*{6. Travelling Index }

We write the \( j \)-th spatial mode as:

\begin{equation}
\underline{\phi}_j = \Re[\underline{\phi}_j] + i \Im[\underline{\phi}_j] = \underline{\phi_j^{\Re}} + i \underline{\phi_j^{\Im}}
\end{equation}

We compute the complex matrix \( \underline{\underline{W}}_j \) and its Gram matrix \( \underline{\underline{G}}_j \) :

\begin{equation}
\underline{\underline{W}}_j(x) = 
\begin{bmatrix}
\underline{\phi_j^{\Re}} & \underline{\phi_j^{\Im}}
\end{bmatrix}
\in \mathbb{R}^{N_x\times2}
\end{equation}

\begin{equation}
\underline{\underline{G}}_j(x) = \underline{\underline{W}}_j^{T} \underline{\underline{W}}_j
\end{equation}

\begin{equation}
\underline{\underline{G}}_j(x) = \begin{bmatrix}
 (\underline{\phi}_j^{\Re})^{T} \underline{\phi}_j^{\Re} &  (\underline{\phi}_j^{\Re})^{T} \underline{\phi}_j^{\Im}  \\
 (\underline{\phi}_j^{\Im})^{T} \underline{\phi}_j^{\Re}  & (\underline{\phi}_j^{\Im})^{T} \underline{\phi}_j^{\Im} 
\end{bmatrix} =
\begin{bmatrix}
a & c \\
c & b
\end{bmatrix}
\end{equation}

Then we define:
\begin{equation}
\alpha_j = \frac{1}{\kappa(W_j)} = \frac{\sigma_{\min}}{\sigma_{\max}} = \sqrt{\frac{\lambda_{min}^{G}}{\lambda_{max}^{G}}}
\end{equation}
where \( \kappa(W) \) is the condition number of the matrix \( W \), and \( \sigma_{\min} \), \( \sigma_{\max} \) are the smallest and largest singular values. In this context, the singular values of \( W_j \) are the square roots of \( G_j \)'s eigenvalues \( \lambda^{G} \) \cite{golub_vanloan_2013}.

We use the quadratic formulas for the eigenvalues being the roots of the characteristic polynomial of the matrix since $G_j$ matrixes has a 2 by 2 size \cite{meyer_2000}. The travelling index \( \alpha_j \) is then computed as :

\begin{equation}
\alpha_j  = 
\sqrt{\frac{a + b - \sqrt{(a - b)^2 + 4c^2}}{a + b + \sqrt{(a - b)^2 + 4c^2}}}
\end{equation}

\begin{itemize}
    \item \( \alpha_j = 1 \): the mode corresponds to a perfectly travelling wave (circular trace in the complex plane).
    \item \( \alpha_j = 0 \): the mode is purely standing (real or imaginary only).
\end{itemize}

This discrete formulation matches the practical implementation used in numerical algorithms and allows direct comparison between theoretical and computational characterizations of wave motion.

\newpage

\newpage
\section{Example n°1 : water waves in a tank}
\label{IV}

As an example, we'll look at the motion of a fluid's free surface—in simple terms, waves. We simulate fluid motion in a rectangular tank of width \( L=400\,\mathrm{mm} \) and water height \( h=100\,\mathrm{mm} \). The surface motion (or wave) is described as the superposition of two standing or travelling wave modes, whose characteristics (amplitude, frequency, wavelength) are derived from the linear Airy wave theory \cite{lamb_hydrodynamics_1945}. In particular, we have the dispersion relation of the waves that link the pulsations and the wavelengths :

\begin{equation}
    \omega_{n}^2 = gk_{n}\tanh (k_nh)
\end{equation}

with, in the case of sloshing  the wavenumber being \cite{lamb_hydrodynamics_1945}:

\begin{equation}
    k_n = \frac{2 \pi}{\lambda_n} = \frac{n\pi}{L} 
\end{equation}

With this theoretical case, we aim at finding back what would give a 2D spatio-temporal FFT since here both space and time can be decomposed onto Fourier basis.

\subsection*{1. Signal Construction}

The signal \( s(x,t) \) is defined as:

\begin{equation}
s(x,t) = s_1(x,t) + s_2(x,t)
\end{equation}
with:
\begin{equation}
s_i(x,t) = A_i \left[ \sin(\omega_i t) \sin\left( \frac{2\pi x}{\lambda_i} \right) + \alpha_i \cos(\omega_i t) \cos\left( \frac{2\pi x}{\lambda_i} \right) \right]
\end{equation}

where:
\begin{itemize}
  \item $x \in [-\frac{L}{2},\frac{L}{2}]$, $t \in [0,T]$.
  \item \( \lambda_i \) are the wavelengths, \( \omega_i \) the angular frequencies.
  \item \( \alpha_1 \in [-1,1]\) controls whether the first wave is standing or travelling. Here we only modify this parameter for the first mode. We will demonstrate that it is the travelling index of the first mode.
  \item \( A_i \) are amplitudes.
\end{itemize}

The signal is discretized in time and space, forming the matrix \( \mathbf{S} \) of size \( N_t \times N_x \), with time samples along the rows and spatial samples along the columns.

We use numerically for this example:\begin{itemize}
    \item $L=400$ mm, $A_1 = 15$, $A_2 = 4$ (amplitudes in arbitrary units).
    \item Mode $n_1 = 1$ and $n_2=3$ (sloshing modes).
    \item $\alpha_1 = \alpha_2 = 0$ (both components are stationnary)
    \item $N_t = 1000$, $N_x = 250$.
\end{itemize}

\textbf{This example is derived in Python in} \textit{test\_COD\_ex\_1.py}.

\newpage

\begin{figure}[ht!]
    \centering
    \includegraphics[scale=0.55]{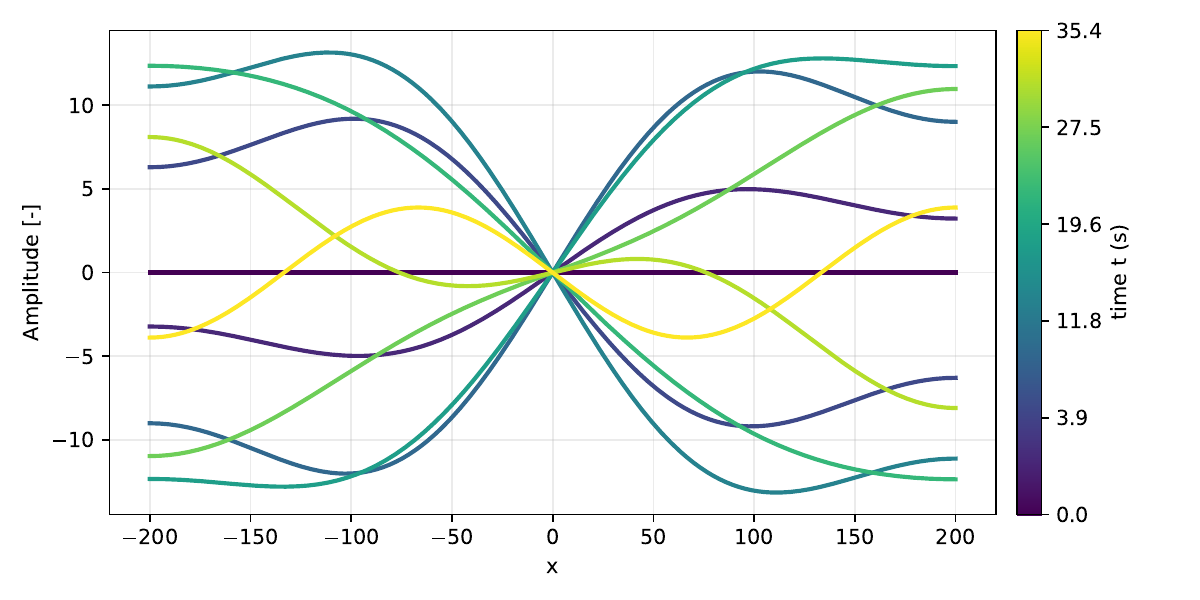}
    \caption[]{Spatio-temporal visualization of the surface waves}
    \label{fig2}
\end{figure}

            \begin{figure}[ht!]
                \begin{minipage}[c]{.46\linewidth}
     \centering
    \includegraphics[scale=0.55]{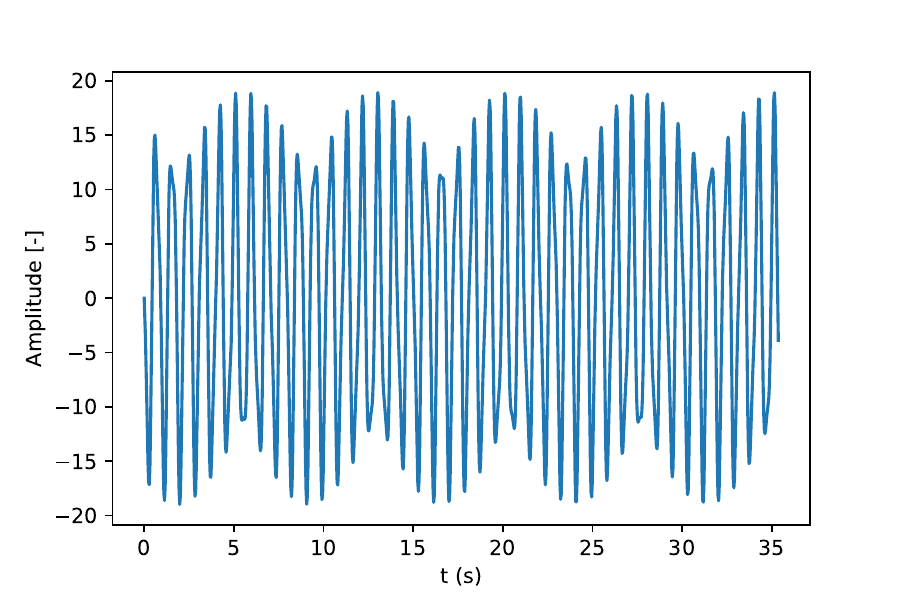}
                \end{minipage}
                \hfill%
                \begin{minipage}[c]{.46\linewidth}
                    \centering
                    \vspace{-0.1mm}  
                    \includegraphics[scale =0.55]{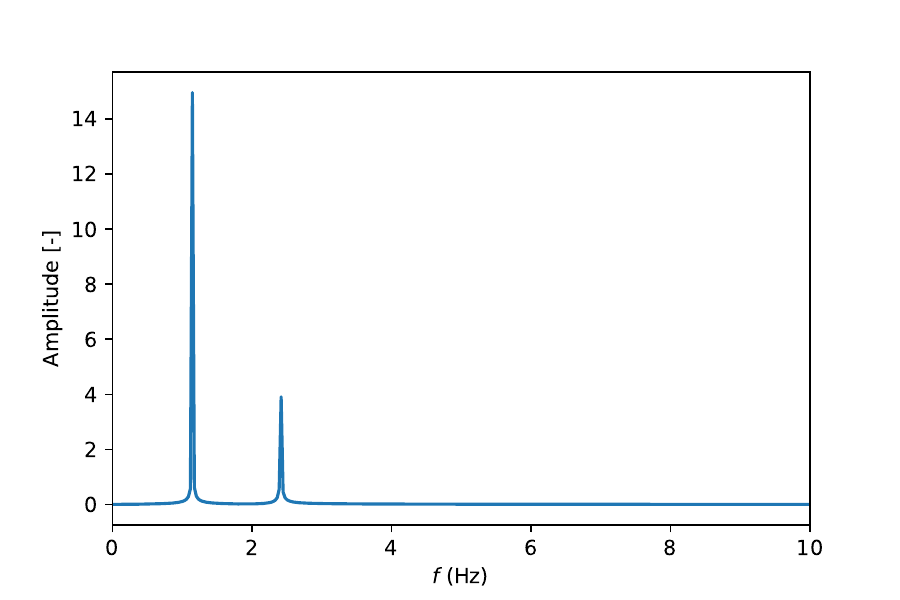}

                \end{minipage}
            \caption[]{Time signal (\textit{left}) and Fourier spectrum (\textit{right}) at $x=-200$ mm (on the wall of the tank, antinode of oscillations for both modes)}
                \label{Fig3}
            \end{figure}

We clearly see on Figure \ref{Fig3} the two components of the signal in the Fourier spectrum, plus being at the antinode allows to also verify that the amplitudes for both modes are right. C.O.D. will help us to decouple these two components thanks to spatial orthogonality. In this example, the calculus can be done analytically.

\subsection*{2. Some theoretical insights}

We consider only $s_1(x,t)$ to show the general idea of C.O.D., because here $s_1$ and $s_2$ are orthogonal and well separated in frequency, which implies linearity of the C.O.D.

\begin{equation}
s_1(x, t) = A_1 \left[ \sin(\omega_1 t) \sin\left( \frac{2\pi x}{\lambda_1} \right) + \alpha_1 \cos(\omega_1 t) \cos\left( \frac{2\pi x}{\lambda_1} \right) \right]
\end{equation}

This can be recognized as a combination that forms a travelling wave when $\alpha_1 = 1$ or $-1$ and a standing wave when $\alpha_1 =0$. Our goal is to analyze the spatial structure after applying the Hilbert transform in time and relate the resulting complex field to the travelling index. Doing so, we will also compute the complex spatial component $\phi_1(x)$ and the complex temporal coefficient $a_1(t)$ so that :

\begin{equation}
s_1(x, t) = \Re[a_1(t)\phi_1(x)]
\end{equation}

To achieve the decomposition, one need to compute the Hilbert transform \(\widetilde{s}(x,t)\) with respect to time \(t\) to obtain the complex signal:
\begin{equation}
s_{c,1}(x,t) = s_1(x,t) + i\, \widetilde{s_1}(x,t)
\end{equation}

Recall the Hilbert transform of sinusoids in time:
\begin{equation}
\mathcal{H}[\sin(\omega t)] = -\cos(\omega t), \quad \mathcal{H}[\cos(\omega t)] = \sin(\omega t)
\end{equation}

Applying linearity, we get:
\begin{equation}
\widetilde{s_1}(x,t) = -\cos(\omega_1 t)  \sin\left( \frac{2\pi x}{\lambda_1} \right) + \alpha_1 \sin(\omega_1 t) \cos\left( \frac{2\pi x}{\lambda_1} \right)
\end{equation}

Therefore, the complex analytic signal is:

\begin{equation}
\begin{aligned}
s_{c,1}(x,t) &= s_1(x,t) + i \widetilde{s_1}(x,t) \\
&= A_1 \left[ \sin(\omega_1 t) \sin\left( \frac{2\pi x}{\lambda_1} \right) + \alpha_1 \cos(\omega_1 t) \cos\left( \frac{2\pi x}{\lambda_1} \right) \right] \\
&\quad + i A_1 \left[ -\cos(\omega_1 t) \sin\left( \frac{2\pi x}{\lambda_1} \right) + \alpha_1 \sin(\omega_1 t) \cos\left( \frac{2\pi x}{\lambda_1} \right) \right]
\end{aligned}
\end{equation}

Group terms by \( \sin(\omega_1 t) \) and \( \cos(\omega_1 t) \):

\begin{equation}
s_{c,1}(x,t) = A_1 \left[ \sin(\omega_1 t) \bigg( \sin\left( \frac{2\pi x}{\lambda_1} \right) + i \alpha_1 \cos\left( \frac{2\pi x}{\lambda_1} \right) \bigg) + \cos(\omega_1 t) \bigg( \alpha_1 \cos\left( \frac{2\pi x}{\lambda_1} \right) - i \sin\left( \frac{2\pi x}{\lambda_1} \right) \bigg) \right]
\end{equation}

Using Euler's formula for the time part:

\begin{equation}
e^{i \omega_1 t} = \cos(\omega_1 t) + i \sin(\omega_1 t)
\end{equation}

we can rewrite the complex analytic signal as:

\begin{equation}
s_{c,1}(x,t)) = A_1 e^{i \omega_1 t} \left[ \sin\left( \frac{2\pi x}{\lambda_1} \right) - i \alpha_1 \cos\left( \frac{2\pi x}{\lambda_1} \right) \right]
\end{equation}

Therefore,

\begin{equation}
s_{c,1}(x,t) = a_1(t) \phi_1(x),
\end{equation}

with

\begin{equation}
a_1(t) = A_1 e^{i \omega_1 t}, \quad \phi_1(x) = \sin\left( \frac{2\pi x}{\lambda_1} \right) - i \alpha_1 \cos\left( \frac{2\pi x}{\lambda_1} \right).
\end{equation}

This confirms the decomposition of the complex analytic signal into a temporal coefficient and spatial mode encoding the travelling behavior through \(\alpha_1\).

We of course have :

\begin{equation}
s_{1}(x,t) = \Re(a_1(t) \phi_1(x))
\end{equation}

Reasoning by linearity will allow to do this also for $s_2$ and therefore construct the full complex orthogonal decomposition of $s$.

Let's see how the discrete numerical calculus gives in terms of decomposition. First we look at on the next figures the energies of each components, $\lambda^{e}_{1}$ and $\lambda^{e}_{2}$, the amplitudes $A_1$ and $A_2$. 

\newpage

            \begin{figure}[ht!]
                \begin{minipage}[c]{.46\linewidth}
     \centering
    \includegraphics[scale=0.55]{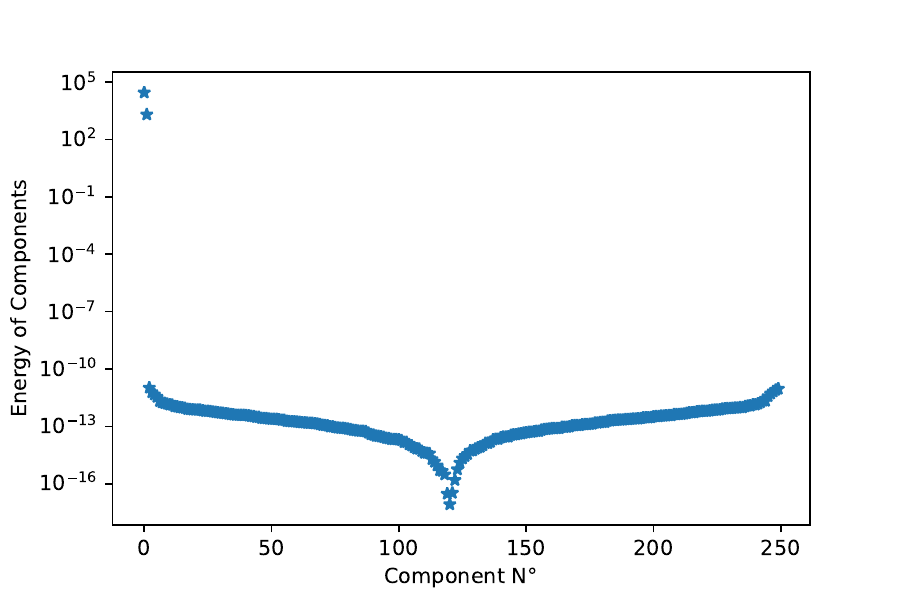}
                \end{minipage}
                \hfill%
                \begin{minipage}[c]{.46\linewidth}
                    \centering
                    \vspace{-0.1mm}  
                    \includegraphics[scale =0.55]{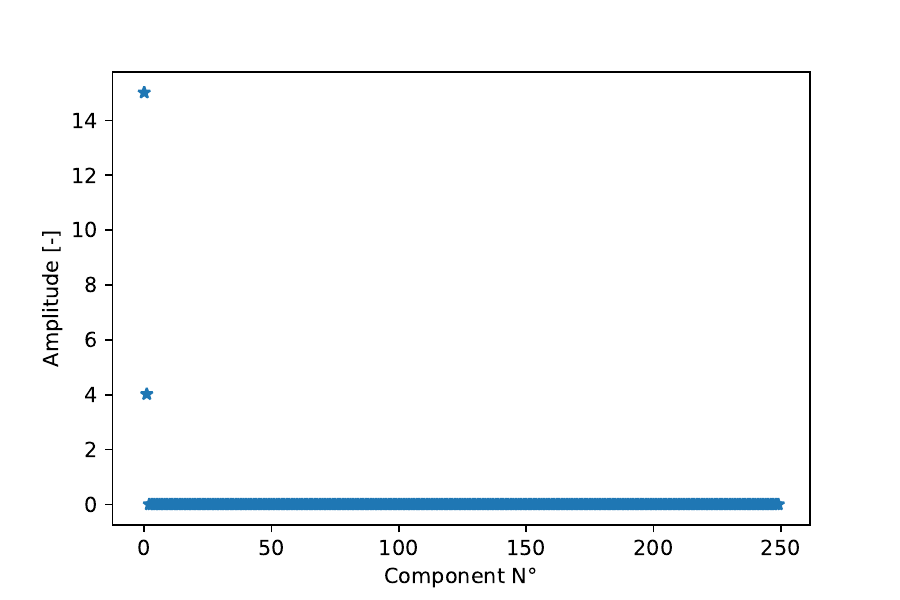}

                \end{minipage}
            \caption[]{Energy of the components $\lambda^{e}_{i}$ (\textit{left}) and corresponding oscillations amplitudes $A_i$ (\textit{right})}
            \label{fig4}
            \end{figure}

\begin{figure}[ht!]
                \begin{minipage}[c]{.46\linewidth}
     \centering
    \includegraphics[scale=0.55]{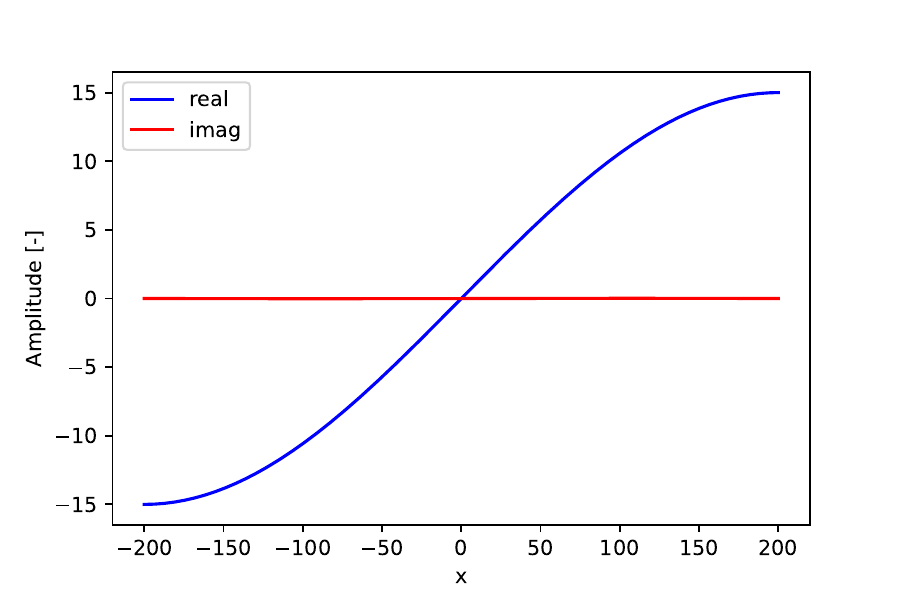}
                \end{minipage}
                \hfill%
                \begin{minipage}[c]{.46\linewidth}
                    \centering
                    \includegraphics[scale =0.55]{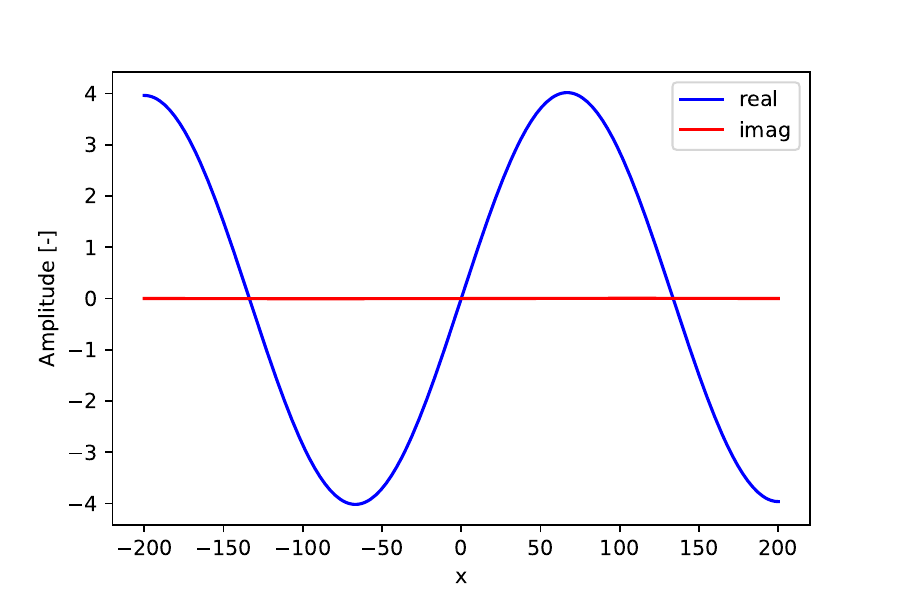}
                \end{minipage}
            \caption[] {$\phi_i^{\Re}(x)$ and $\phi^{\Im}_i(x)$ for $i=1$ (\textit{left}) and $i=2$ (\textit{right})}
            \label{fig5}
            \end{figure}

\begin{figure}[ht!]
    \centering
    \includegraphics[scale=0.65]{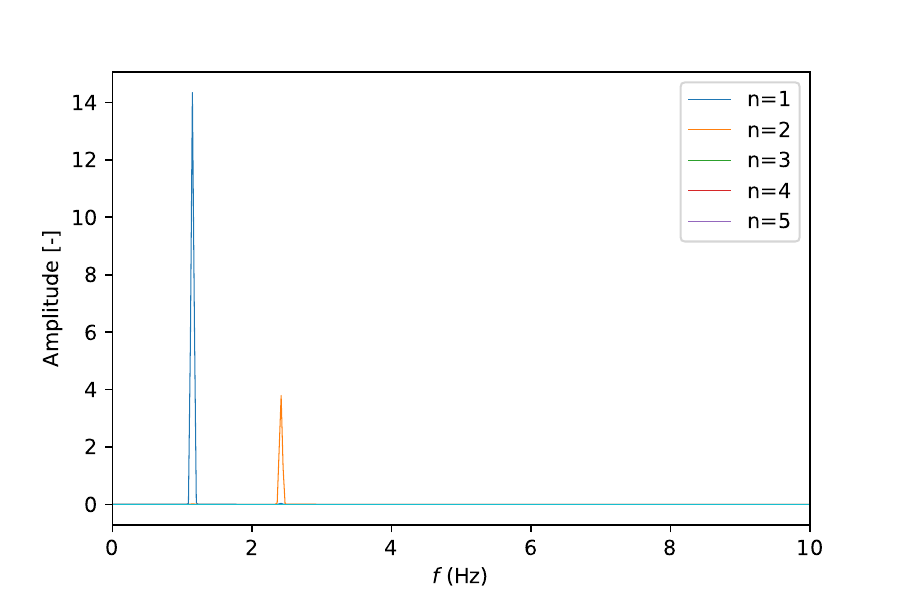}
    \caption[]{Fourier spectra of the 5 first components of the C.O.D.}
    \label{fig6}
\end{figure}

Figure \ref{fig4} shows that the two orthogonal modes are well separated : except the two wanted components, all the quantities are close to 0. The right figures shows that the C.O.D. is finding back the values of amplitudes ($A_1 = 4$ and $A_2=15$) with precision. We can also look at the spatial forms retrieved by the C.O.D on Figure \ref{fig5}, as well as the time Fourier transform of each signal $a_i(t)$ on Figure \ref{fig6}. We find back the theoretical predictions.

\newpage

\textbf{Travelling Index :}

We have as the spatial components of the component n°1 (same analysis can be done for component n°2):
\begin{equation}
\phi_j^{\Re}(x) = \sin\!\Bigl(\frac{2\pi x}{\lambda_1}\Bigr), 
\quad
\phi_j^{\Im}(x) = -\alpha_1 \cos\!\Bigl(\frac{2\pi x}{\lambda_1}\Bigr)
\end{equation}

We compute then the Gram matrix, being

\begin{equation}G_j(x) = \begin{bmatrix}
\langle \phi_j^{\Re}(x), \phi_j^{\Re}(x) \rangle & \langle \phi_j^{\Re}(x), \phi_j^{\Im}(x) \rangle \\
\langle \phi_j^{\Im}(x), \phi_j^{\Re}(x) \rangle & \langle \phi_j^{\Im}(x), \phi_j^{\Im}(x) \rangle
\end{bmatrix} =
\begin{bmatrix}
a & c \\
c & b
\end{bmatrix}
\end{equation}

Over one spatial period  $\lambda_1$, compute
\begin{equation}
a = \int_{0}^{\lambda_1} \sin^2\!\Bigl(\tfrac{2\pi x}{\lambda_1}\Bigr)\,dx
= \frac{\lambda_1}{2},
\end{equation}
\begin{equation}
b = \int_{0}^{\lambda_1} \bigl(\alpha_1 \cos\tfrac{2\pi x}{\lambda_1}\bigr)^{2}\,dx
= \alpha_1^2 \,\frac{\lambda_1}{2},
\end{equation}
\begin{equation}
c = \int_{0}^{\lambda_1} -\sin\!\Bigl(\tfrac{2\pi x}{\lambda_1}\Bigr)\,\alpha_1 \cos\!\Bigl(\tfrac{2\pi x}{\lambda_1}\Bigr)\,dx
=  -\alpha_1 \int_{0}^{\lambda_1} \tfrac12\sin\!\Bigl(\tfrac{4\pi x}{\lambda_1}\Bigr)\,dx
= 0.
\end{equation}

Thus the Gram matrix eigenvalues are $a$ and $b$, but we can verify that the formula given in the precedent section works well :
\begin{equation}
\lambda_{\pm}^G
=\frac{a+b}{2}\pm\frac12\sqrt{(a-b)^2+4c^2}
=\frac{\tfrac{\lambda_1}{2}+\alpha_1^2\tfrac{\lambda_1}{2}}{2}
\pm\frac12\bigl|\tfrac{\lambda_1}{2}-\alpha_1^2\tfrac{\lambda_1}{2}\bigr|.
\end{equation}
Since \(-1\le\alpha_1\le1\), we have  \(a\le b\), and this becomes :
\begin{equation}
\lambda_{\max}^G = a = \frac{\lambda_1}{2}, 
\quad
\lambda_{\min}^G = b = \frac{\alpha_1^2\lambda_1}{2}.
\end{equation}

Hence the travelling index is $\alpha$ :
\begin{equation}
\alpha
= \sqrt{\frac{\lambda_{\min}^G}{\lambda_{\max}^G}}
= \sqrt{\frac{\alpha_1^2\,\tfrac{\lambda_1}{2}}{\tfrac{\lambda_1}{2}}}
= |\alpha_1|.
\end{equation}

We demonstrated that $\alpha_1$ is the travelling index of associated with $s_1$, in absolute value.

In the example, we took $\alpha_1=0$. The numerics given by the test case \textit{test\_COD\_ex\_1.py} give $\alpha_1 \approx 8.2 \times 10^{-4}$, which is a good approximate value given the refined mesh we took.

\newpage
\section{Example n°2 : temporally decreasing standing wave}
\label{VI}

For this example, we imagine a standing wave that is progressively fading as time go. One can imagine that the standing wave is generated at $t=0$ and then dissipation is damping it uniformly in space. Here this example demonstrate that in the case of non-purely oscillatory motion, we still recover with good precision the modes, especially if the space form is well-defined. 

\subsection*{1. Signal construction}

We consider the damped sinusoidal signal:
\begin{equation}
s_1(x,t) = A_1\,e^{-\gamma t}\,\sin(\omega_1 t)\,\sin\left(\frac{2\pi x}{\lambda_1}\right)
\end{equation}

We use numerically for this example:\begin{itemize}
    \item $L=400$ mm, $\lambda_1 = 300$ mm.
    \item $A_1 = 16$ (amplitudes in arbitrary units).
    \item $\omega_1 = 2\pi f_1$, $~~f_1 = 5$ Hz.
    \item $\alpha_1 = 0$.
    \item $\gamma = 1$ s$^{-1}$ (damping).
    \item $N_t = 500$, $N_x = 1200$.
\end{itemize}

\textbf{This example is derived in Python in} \textit{test\_COD\_ex\_3.py}.

\begin{figure}[ht!]
    \centering
    \includegraphics[scale=0.7]{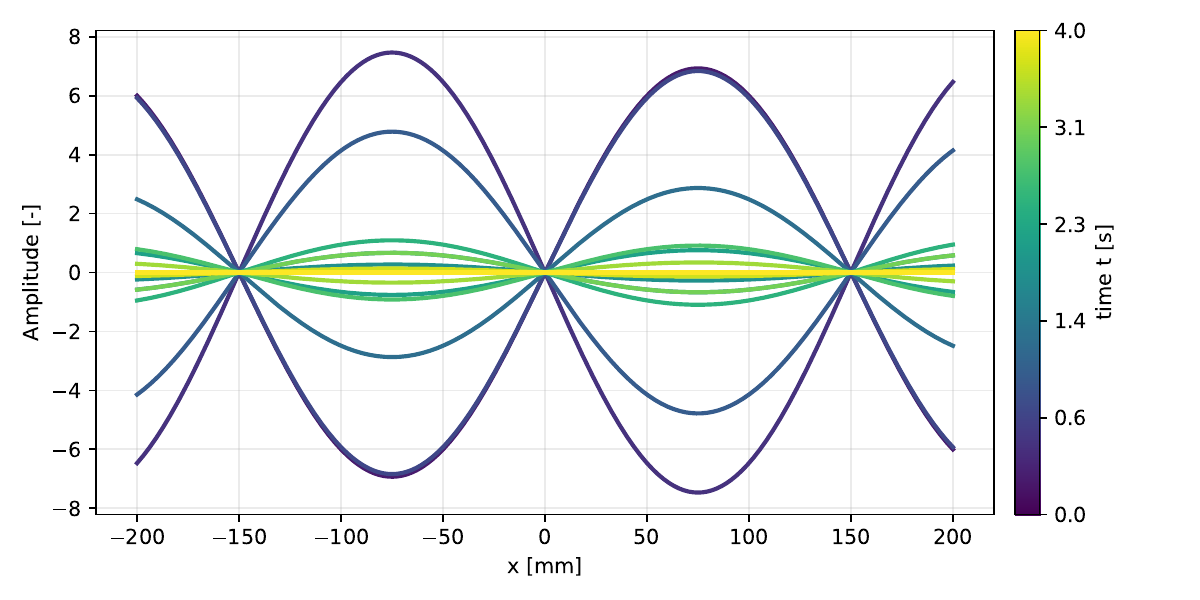}
    \caption[]{Spatio-temporal visualization of the damped waves}
    \label{fig_opm}
\end{figure}

\begin{figure}[ht!]
                \begin{minipage}[c]{.46\linewidth}
     \centering
    \includegraphics[scale=0.55]{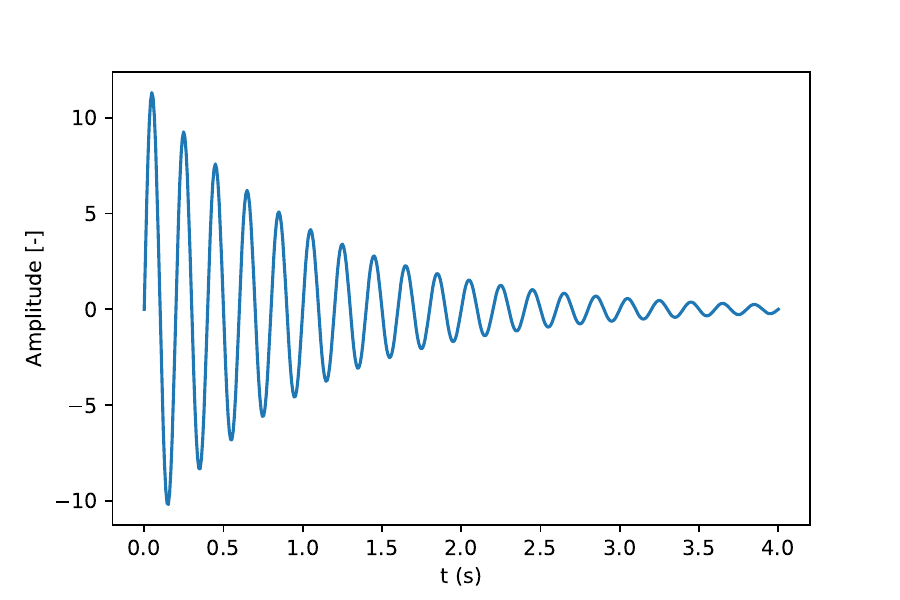}
                \end{minipage}
                \hfill%
                \begin{minipage}[c]{.46\linewidth}
                    \centering
                    \vspace{-.1mm}  
                    \includegraphics[scale =0.55]{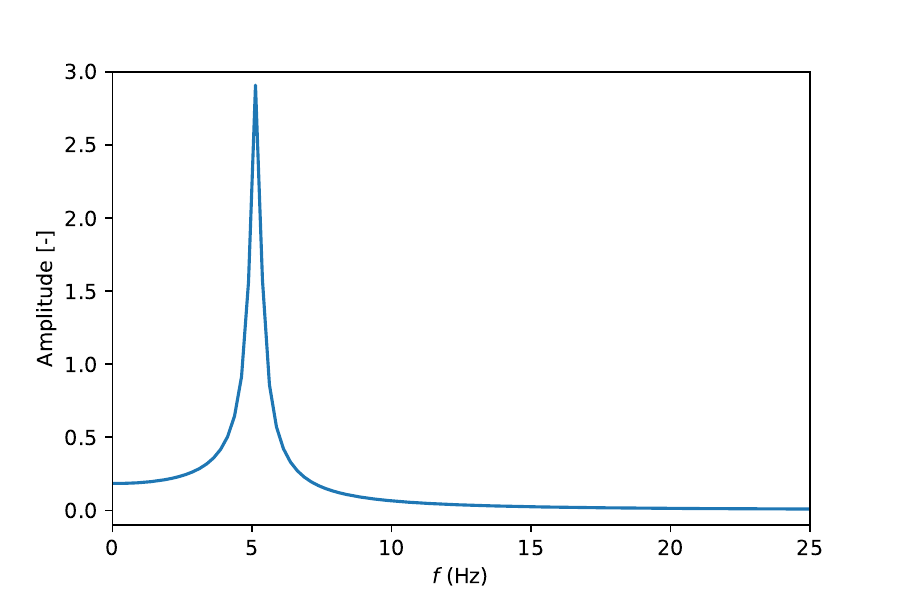}

                \end{minipage}
            \caption[]{Time signal (\textit{left}) and Fourier spectrum (\textit{right}) at $x=10$ mm}
                \label{Fig_iop}
            \end{figure}

We clearly see on Figure \ref{fig_opm} that the spatial form is being constant over time (a sinusoidal standing wave), but that the amplitude is damped over time. To convince ourselves of the exponential decay of the signal, we can look at Figure \ref{Fig_iop}. Indeed, both the temporal signal and its time Fourier transform are consistent with a exponential decay.

\newpage

\subsection*{2. Some theoretical insights}

We compute the analytic signal using the Hilbert transform in time:
\begin{equation}
s_{c,1}(x,t) = s_1(x,t) + i\,\widetilde{s_1}(x,t)
\end{equation}

Since the Hilbert transform is not multiplicative, we must treat the damping carefully. In general:
\begin{equation}
\mathcal{H}[e^{-\gamma t} \sin(\omega t)] \neq e^{-\gamma t} \cdot \mathcal{H}[\sin(\omega t)]
\end{equation}

However, for "slow damping" (i.e., \(\gamma \ll \omega\)), we approximate:
\begin{equation}
\mathcal{H}[e^{-\gamma t} \sin(\omega t)] \approx -e^{-\gamma t} \cos(\omega t)
\end{equation}

The approximation is justified at the end of this section. Thus, we get:
\begin{equation}
\widetilde{s_1}(x,t) \approx -A_1\,e^{-\gamma t}\,\cos(\omega_1 t)\,\sin\left(\frac{2\pi x}{\lambda_1}\right)
\end{equation}

So the complexified analytic signal is:
\begin{equation}
s_{c,1}(x,t) \approx A_1\,e^{-\gamma t} \left[ \sin(\omega_1 t) - i \cos(\omega_1 t) \right] \sin\left(\frac{2\pi x}{\lambda_1} \right)
\end{equation}

Recognizing:
\begin{equation}
\sin(\omega t) - i \cos(\omega t) = -i e^{i \omega t}
\end{equation}
we obtain:
\begin{equation}
s_{c,1}(x,t) \approx -i\,A_1\,e^{-\gamma t}\,e^{i\omega_1 t}\,\sin\left(\frac{2\pi x}{\lambda_1} \right)
\end{equation}

We identify:
\begin{equation}
s_{c,1}(x,t) = a_1(t) \cdot \phi_1(x)
\quad \text{with} \quad
a_1(t) = -i\,A_1\,e^{-\gamma t}\,e^{i\omega_1 t}, \quad \phi_1(x) = \sin\left(\frac{2\pi x}{\lambda_1} \right)
\end{equation}

Therefore:
\begin{equation}
s_1(x,t) = \Re[a_1(t) \cdot \phi_1(x)]
\end{equation}

We can directly calculate the travelling index. The spatial mode is real:
\begin{equation}
\phi_1^{\Re}(x) = \sin\left(\frac{2\pi x}{\lambda_1} \right), \quad \phi_1^{\Im}(x) = 0
\end{equation}

Defining the Gram matrix:
\begin{equation}
G = 
\begin{bmatrix}
\langle \phi_1^{\Re}, \phi_1^{\Re} \rangle & \langle \phi_1^{\Re}, \phi_1^{\Im} \rangle \\
\langle \phi_1^{\Im}, \phi_1^{\Re} \rangle & \langle \phi_1^{\Im}, \phi_1^{\Im} \rangle
\end{bmatrix}
=
\begin{bmatrix}
a & 0 \\
0 & 0
\end{bmatrix}
\end{equation}

The eigenvalues of \(G\) are \(\lambda_1^{G} = a\), \(\lambda_2^{G} = 0\), so no matter what $a$ is equal to:
\begin{equation}
\alpha_1 = \sqrt{\frac{\lambda_{\min}^{G}}{\lambda_{\max}^{G}}} = 0
\end{equation}

The mode is purely standing, with travelling index:
\begin{equation}
\alpha_1 = 0
\end{equation}

Now that we have done the calculus analytically, let's see how the discrete numerical calculus gives in terms of decomposition. First we look at on the next figures the energies of each components and the amplitudes.

            \begin{figure}[ht!]
                \begin{minipage}[c]{.46\linewidth}
     \centering
    \includegraphics[scale=0.55]{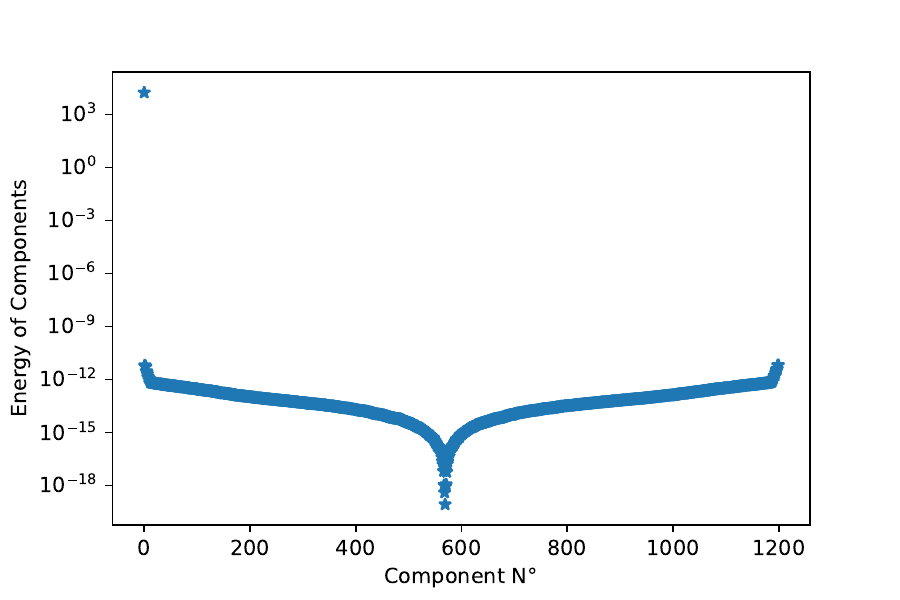}
                \end{minipage}
                \hfill%
                \begin{minipage}[c]{.46\linewidth}
                    \centering
                    \includegraphics[scale =0.55]{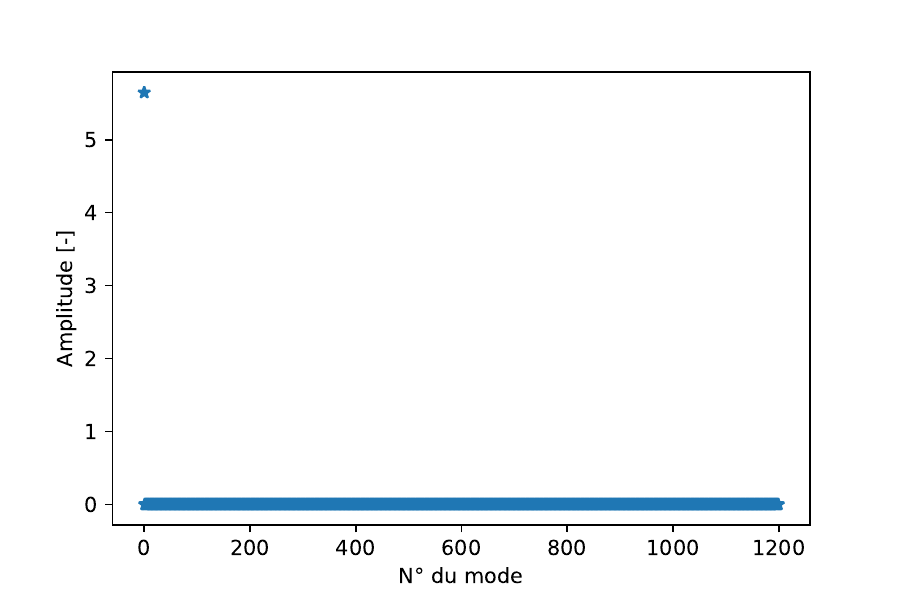}

                \end{minipage}
            \caption[]{Energy of the components $\lambda^{e}_{i}$ (\textit{left}) and corresponding oscillations amplitudes $A_i$ (\textit{right})}
            \label{fig_sap}
            \end{figure}

We recover only one component with all the energy of the signal and this component has an amplitude of nearly 6. Here a verification calculus could be performed using 

\begin{equation}
\lambda_j^1 = \frac{1}{T} \int_0^T |a_1(t)|^2 \, \mathd t.
\end{equation}

and link the energy to the initial value of the signal and the total time since it is an integrated quantity. Here there is no clear definition of the amplitude and this calculus can be tedious, so it is left aside. The main element here is that all other components are zero. 

        \begin{figure}[ht!]
    \centering
    \includegraphics[scale=0.57]{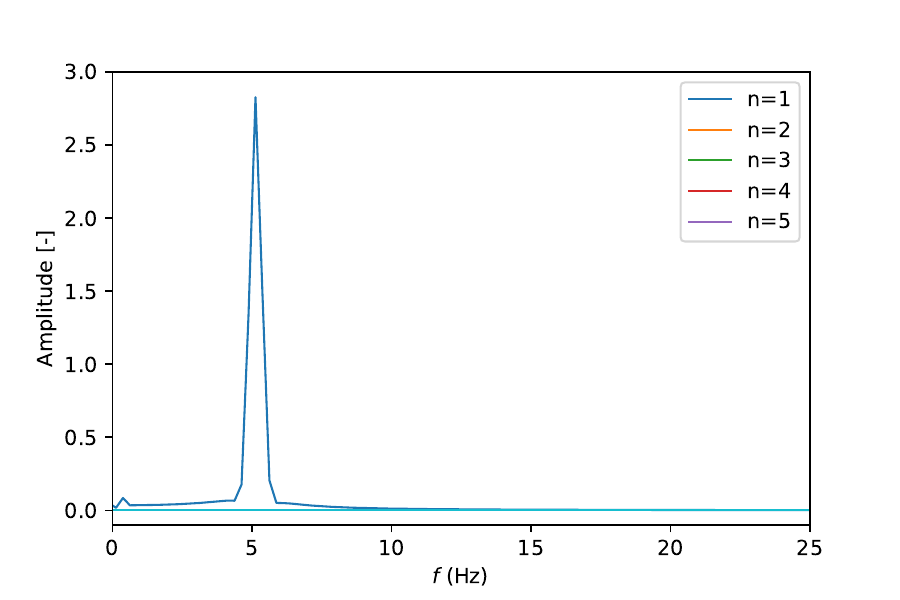}
    \caption[]{Fourier spectra of the 5 first components of the C.O.D.}
    \label{poi}
\end{figure}

\newpage

Nonetheless, we can compare the spectra by displaying the time Fourier spectra of each component of Figure \ref{poi}. The spectrum is not representing as closely as it could the real spectrum on Figure \ref{Fig_iop}. Indeed, instead of having a large peak, this one is closer to only one harmonic, even if it does present a larger frequency band compared to a pure harmonic as seen in example n°1 or n°2. This error is due to the approximation done in the Hilbert transform (see end of this section).

\begin{figure}[ht!]
    \centering
    \includegraphics[scale=0.7]{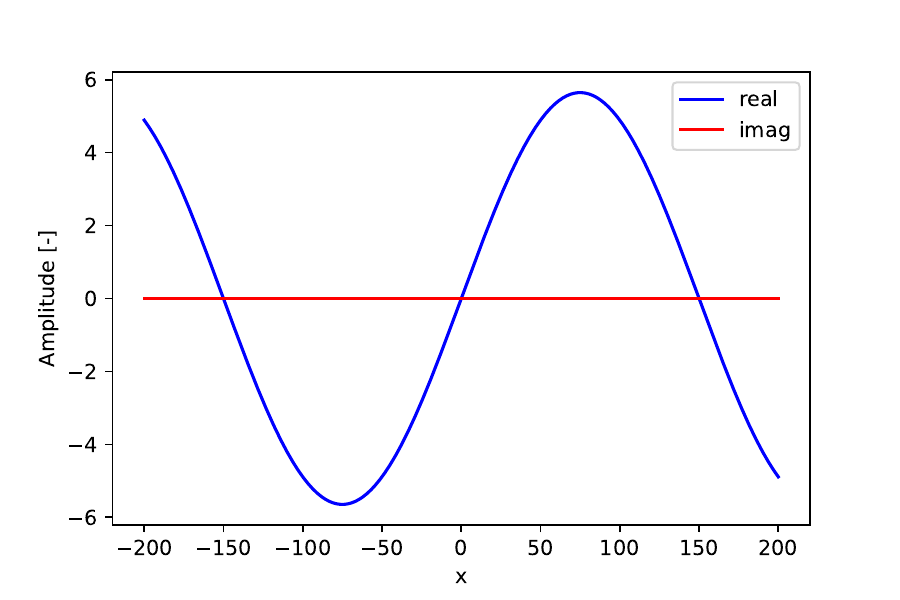}
    \caption[]{$\phi_i^{\Re}(x)$ and $\phi^{\Im}_i(x)$ for $i=1$}
    \label{rtyu}
\end{figure}

We also look at the spatial form of this component on Figure \ref{rtyu}.

We indeed recover $
\phi_1^{\Re}(x) = \sin\left(\frac{2\pi x}{\lambda_1} \right)$ and $\phi_1^{\Im}(x) = 0 $.

\vspace{1cm}

\textbf{Justification of the Approximation :}

Let us consider the general form:
\begin{equation}
f(t) = e^{-\gamma t} \sin(\omega t)
\end{equation}

The Hilbert transform is:
\begin{equation}
\mathcal{H}[f](t) = \frac{1}{\pi} \, \text{p.v.} \int_{-\infty}^\infty \frac{f(\tau)}{t-\tau} \, d\tau
\end{equation}

Although the Hilbert transform is linear, it is not multiplicative in general. However, when \(\gamma \ll \omega\), the envelope changes slowly compared to the oscillation, leading to the approximation:
\begin{equation}
\mathcal{H}[e^{-\gamma t} \sin(\omega t)] \approx - e^{-\gamma t} \cos(\omega t)
\end{equation}

This can be further justified by Fourier analysis. The Fourier transform of:
\begin{equation}
f(t) = e^{-\gamma t} \sin(\omega t)\, 
\end{equation}
is:
\begin{equation}
\hat{f}(\xi) = \Im \left\{ \int_0^\infty e^{-(\gamma + i(\xi - \omega))t} dt \right\} = \frac{\xi - \omega}{\gamma^2 + (\xi - \omega)^2}
\end{equation}

This spectrum is peaked near \(\omega\), and for small \(\gamma\), the phase shift induced by the Hilbert transform (multiplication by \(-i 
~\text{sgn}(\xi)\)) yields an approximate time-domain result:
\begin{equation}
\mathcal{H}[f](t) \approx -e^{-\gamma t} \cos(\omega t)
\end{equation}

\newpage

\begin{figure}[ht!]
    \centering
    \includegraphics[scale=0.6]{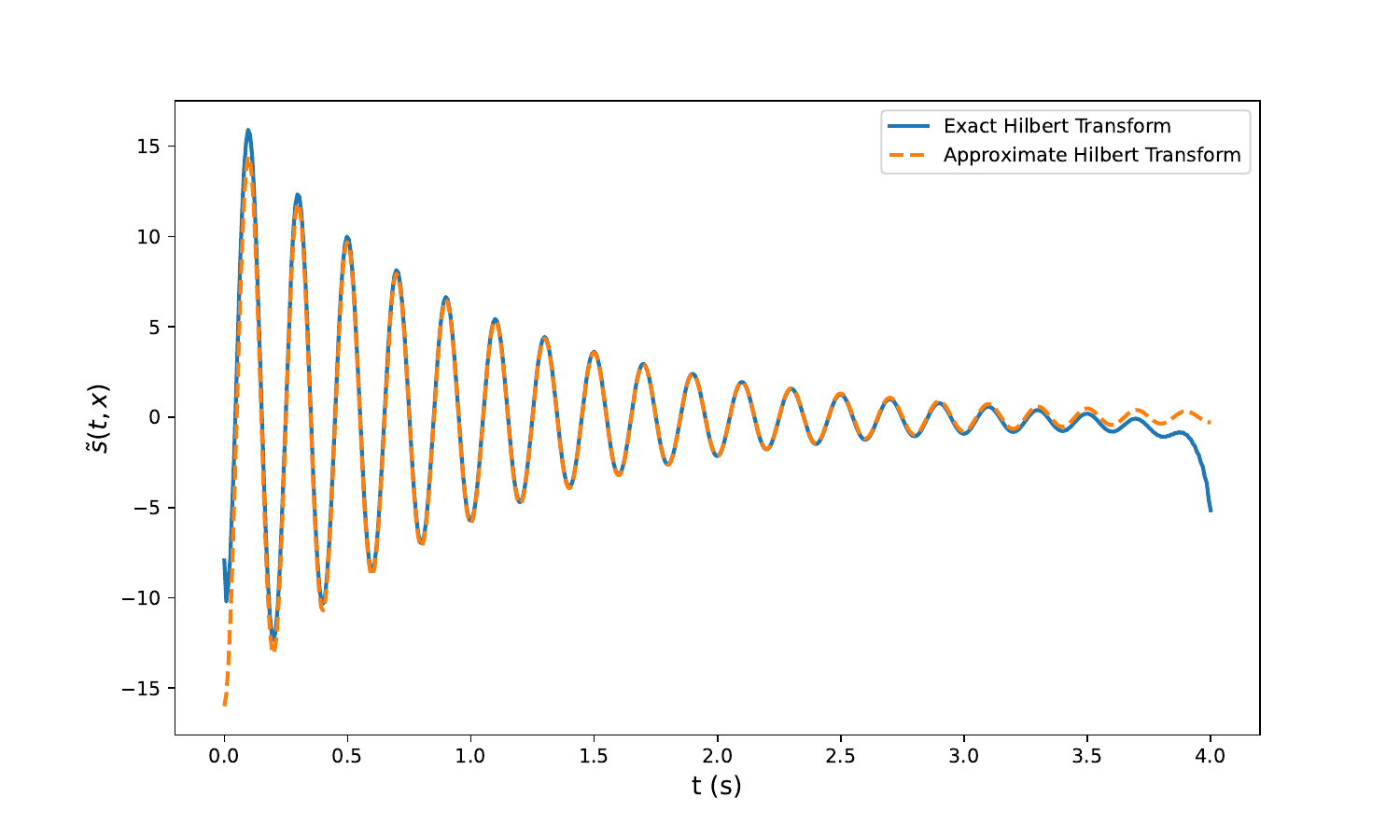}
    \caption[]{Comparison between approximate and exact Hilbert transform}
    \label{Hilbert}
\end{figure}

A visual comparison between the exact Hilbert transform and the approximated one is displayed on Figure \ref{Hilbert}.

\newpage
\section{$~$ Example n°3 : frequency modulated wave}
\label{VII}

This example shows a standing wave where the frequency is modulated over time. The objective is to show that even if a spatial form is oscillating with multiple frequencies, the C.O.D. is able to recover the proper behaviour. 

\subsection*{1. Signal construction}

Consider a signal of the form:
\begin{equation}
s(x, t) = \phi(x) a(t),
\end{equation}
where:
\begin{itemize}
    \item $\phi(x)$ is a fixed real-valued spatial component, here we choose $(0.01 \times x)^3$.
    \item $a(t) = \sin(\theta(t))$ with $\theta(t) = \omega_1 t + \epsilon \sin(\Omega t)$ is a signal with frequency modulation.
\end{itemize}

We use numerically for this example:\begin{itemize}
    \item $L=400$ mm.
    \item $A_1 = 2$ (amplitudes in arbitrary units).
    \item $\omega_1 = 2\pi f_1$, $~~f_1 = 1$ Hz.
    \item $\Omega =  2 \pi F$, $F=0.2$ Hz.
    \item $\alpha_1 = 0$ (travelling index).
    \item $N_t = 1000$, $N_x = 250$.
\end{itemize}

\textbf{This example is derived in Python in} \textit{test\_COD\_ex\_4.py}.

\begin{figure}[ht!]
    \centering
    \includegraphics[scale=0.7]{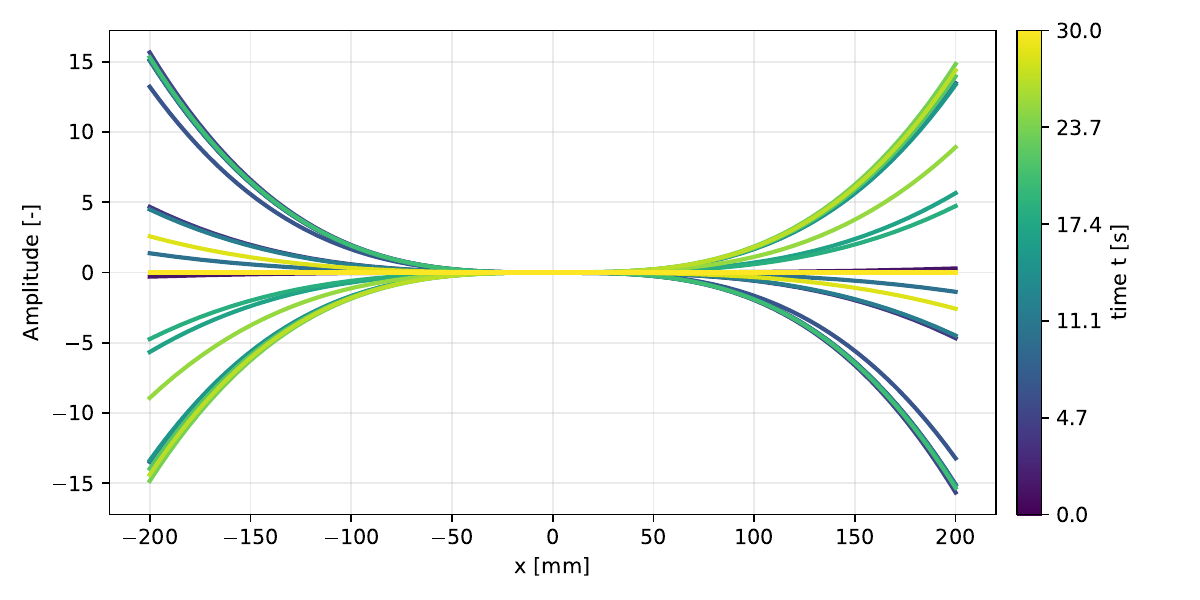}
    \caption[]{Spatio-temporal visualization of the "cubic" waves}
    \label{fig_opm_4}
\end{figure}

\begin{figure}[ht!]
                \begin{minipage}[c]{.46\linewidth}
     \centering
    \includegraphics[scale=0.55]{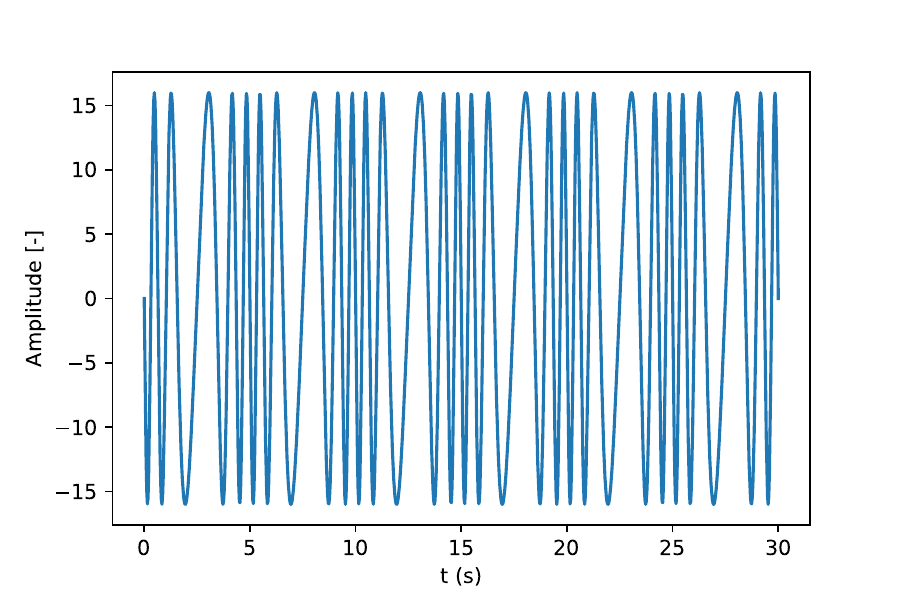}
                \end{minipage}
                \hfill%
                \begin{minipage}[c]{.46\linewidth}
                    \centering
                    \vspace{-.1mm}  
                    \includegraphics[scale =0.55]{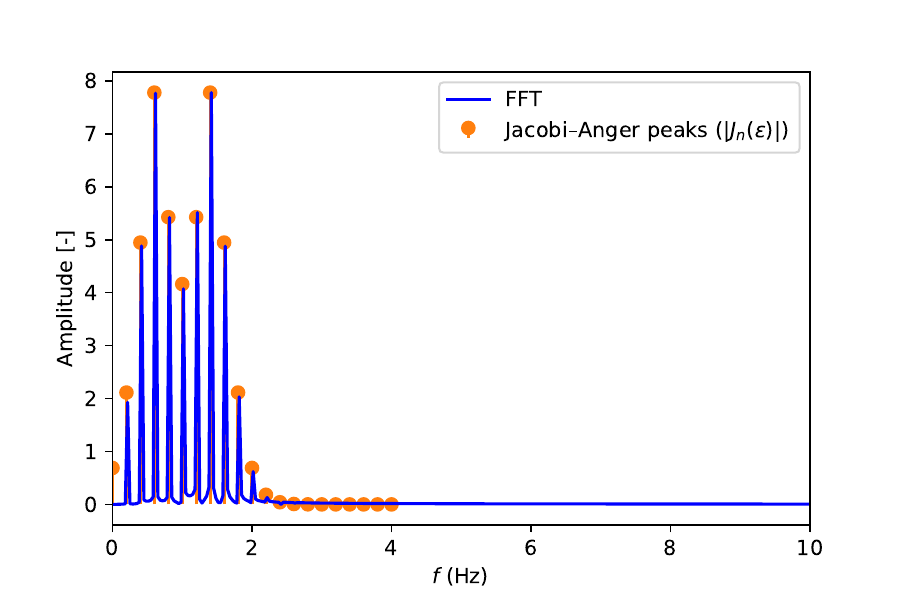}

                \end{minipage}
            \caption[]{Time signal (\textit{left}) and Fourier spectrum (\textit{right}) at $x=-200$ mm}
                \label{Fig_iop_4}
            \end{figure}

All the curves on Figure \ref{fig_opm_4} are effectively of the form $a \times x^3$. It is difficult on these snapshot to see the frequency modulation. For this we focus on Figure \ref{Fig_iop_4}, that displays a time signal at one definite point and its time Fourier transform.

\newpage

We can clearly see on Figure \ref{Fig_iop_4} the frequency modulation. We overlayed on the right plot the theoretical Jacobi-Anger peaks to demonstrate that the frequency modulation in the spectral domain. Let's see how the Complex Orthogonal Decomposition deals with this type of signal. 

\subsection*{2. Some theoretical insights}

We aim to analyze this signal using the \textbf{Complex Orthogonal Decomposition (C.O.D.)}. To do this, we compute the \textbf{analytic signal} using the Hilbert transform:
\begin{equation}
s_c(x,t) = s(x,t) + i\,\widetilde{s}(x,t) = \phi(x) \cdot a_c(t),
\end{equation}
where $a_c(t)$ is the analytic signal of $a(t)$:
\begin{equation}
a_c(t) = a(t) + i\,\mathcal{H}[a](t).
\end{equation}

We write:
\begin{equation}
a(t) = \sin(\omega_1 t + \epsilon \sin(\Omega t)) = \Im\left(e^{i(\omega_1 t + \epsilon \sin(\Omega t))}\right).
\end{equation}

Therefore, the analytic signal is:
\begin{equation}
a_c(t) = e^{i(\omega_1 t + \epsilon \sin(\Omega t))}.
\end{equation}

Using the Jacobi-Anger expansion \cite{abramowitz_stegun_1983}:
\begin{equation}
e^{i \epsilon \sin(\Omega t)} = \sum_{n=-\infty}^{\infty} J_n(\epsilon) e^{in\Omega t},
\end{equation}
where $J_n$ denotes the Bessel function of the first kind of order $n$, we obtain:
\begin{equation}
a_c(t) = e^{i \omega_1 t} \cdot \sum_{n=-\infty}^{\infty} J_n(\epsilon) e^{in\Omega t}
= \sum_{n=-\infty}^{\infty} J_n(\epsilon) e^{i(\omega_1 + n \Omega)t}.
\end{equation}

Thus, the complex signal becomes:
\begin{equation}
s_c(x,t) = \phi(x) \cdot \sum_{n=-\infty}^{\infty} J_n(\epsilon) e^{i(\omega_1 + n \Omega)t}.
\end{equation}

Although $a_c(t)$ contains multiple frequency components, the \textbf{spatial component $\phi(x)$ is identical} for all of them. Hence:
\begin{equation}
s_c(x,t) = \sum_n a_n(t) \cdot \phi(x),
\end{equation}
with $a_n(t) = J_n(\epsilon) e^{i(\omega_1 + n\Omega)t}$.

This means the signal is \textbf{separable}, with only a single spatial mode $\phi(x)$. The associated \textbf{Gram matrix} is of rank 1:
\begin{equation}
G = 
\begin{bmatrix}
\langle \phi^{\Re}, \phi^{\Re} \rangle & 0 \\
0 & 0
\end{bmatrix}
\quad \Rightarrow \quad
\alpha = 0.
\end{equation}

To sum up, modulating the temporal component in frequency:
\begin{itemize}
    \item enriches the \emph{temporal spectrum} of the signal,
    \item but does not introduce additional spatial components,
    \item therefore the \textbf{C.O.D. detects a purely standing mode} (travelling index $\alpha = 0$).
\end{itemize}

Now that we have done the calculus analytically, let's see if the discrete numerical calculus effectively does what is expected. First we look at on the next figures the energies of each components and the amplitudes.

            \begin{figure}[ht!]
                \begin{minipage}[c]{.46\linewidth}
     \centering
    \includegraphics[scale=0.55]{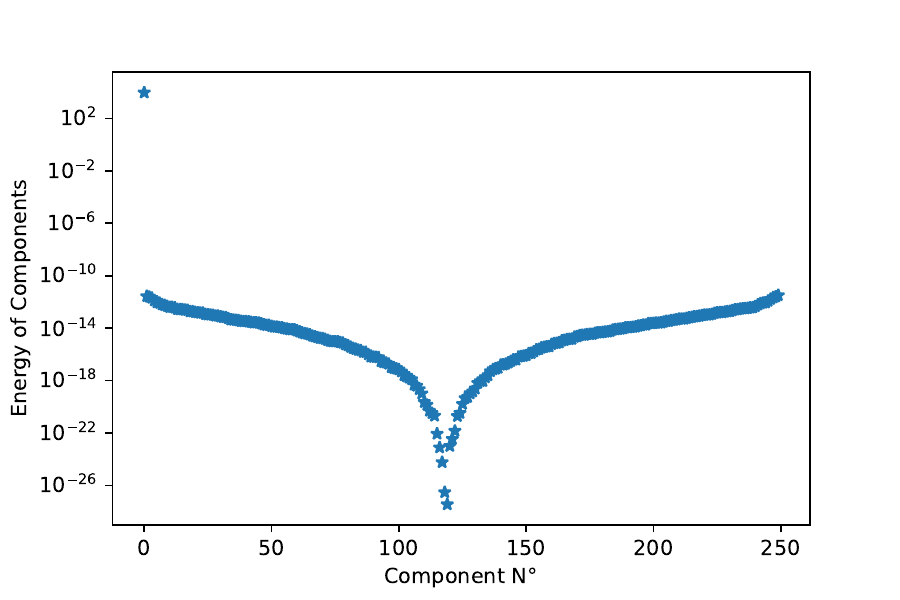}
                \end{minipage}
                \hfill%
                \begin{minipage}[c]{.46\linewidth}
                    \centering
                    \includegraphics[scale =0.55]{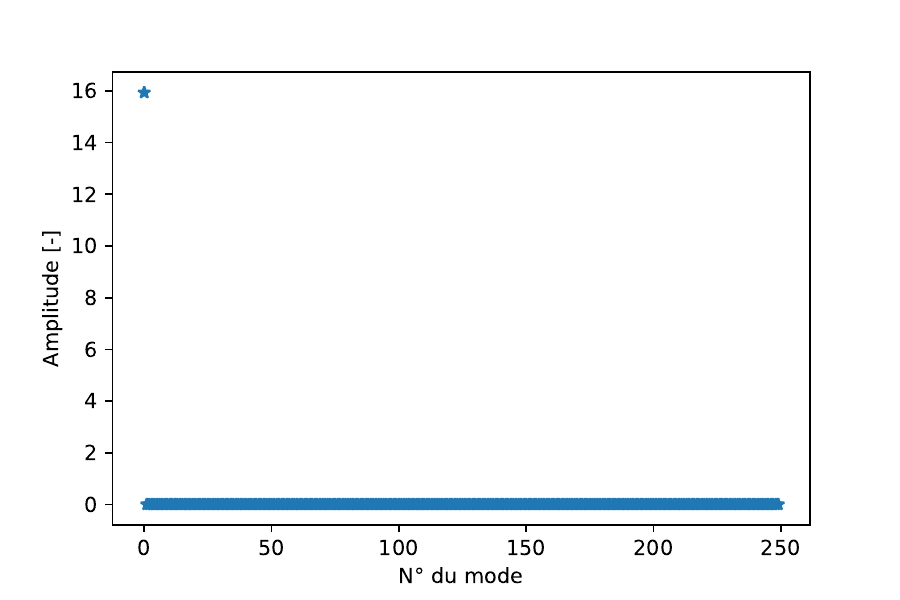}

                \end{minipage}
            \caption[]{Energy of the components $\lambda^{e}_{i}$ (\textit{left}) and corresponding oscillations amplitudes $A_i$ (\textit{right})}
            \label{fig_sap_4}
            \end{figure}

\begin{figure}[ht!]
    \centering
    \includegraphics[scale=0.7]{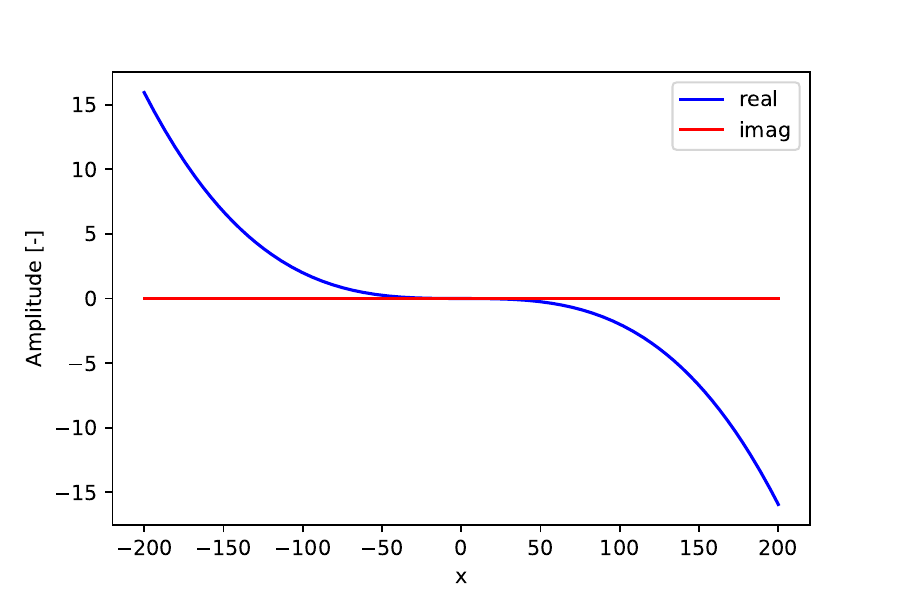}
    \caption[]{$\phi_i^{\Re}(x)$ and $\phi^{\Im}_i(x)$ for $i=1$}
    \label{rtyu_4}
\end{figure}

\begin{figure}[ht!]
    \centering
    \includegraphics[scale=0.8]{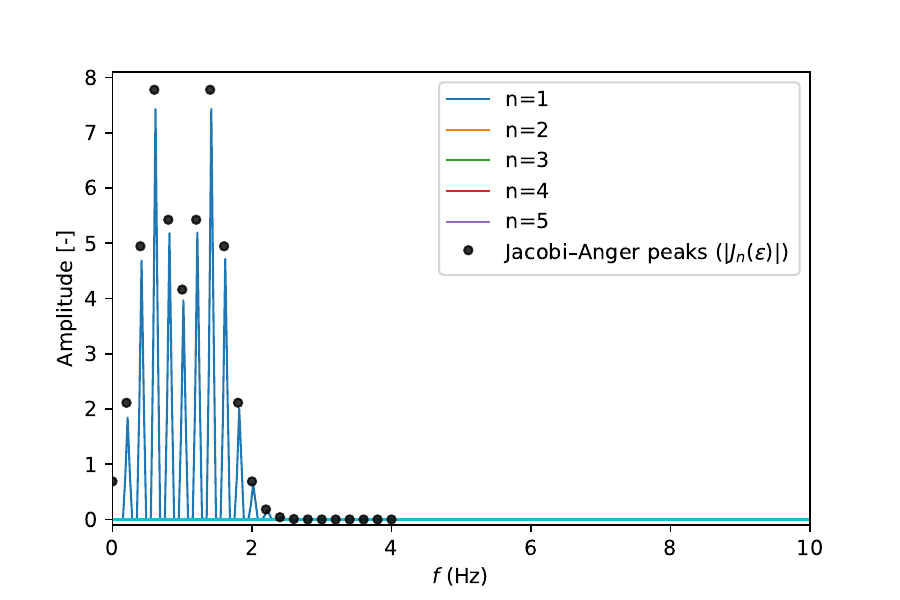}
    \caption[]{Fourier spectra of the 5 first components of the C.O.D.}
    \label{poi_4}
\end{figure}

Energies and amplitudes are effectively recovered, since it is, here again, only one component. Similarly, the cubic form is recovered when looking at the spatial form on Figure \ref{rtyu_4}. More interesting is to look at the time Fourier spectrum of the first five components, on Figure \ref{poi_4}. 

\newpage

We observe here that the first component of the C.O.D gives a very good approximation in term of spectrum. Some energy is lost because of the discretization and the Hilbert transform operation (peaks have underestimated amplitudes) but the frequencies are fairly the same.

In terms of travelling index, the numerical one is $1.25 \times 10^{-17}$, which is also a good approximation of the theoretical value $\alpha_1 = 0$.



\newpage
\section{$~$ Appendice: non-uniform space grid}

In experimental situations, the spatial sampling points are not always evenly spaced. In this part we extend the C.O.D. to non-uniform grids. To do so, we therefore consider a discretely sampled spatio-temporal signal represented by the matrix \( \umat{S} \in \mathbb{R}^{N_t \times N_x} \), where:
\begin{itemize}
    \item \( N_t \) is the number of time steps,
    \item \( N_x \) is the number of spatial points,
    \item each row of \( \umat{S} \) still corresponds to a fixed time \( t_n \),
    \item each column still corresponds to a fixed spatial location \( x_j \), but the grid \( \{x_j\}_{j=1}^{N_x} \) is not necessarily uniform.
\end{itemize}

To correctly approximate continuous inner products and energy integrals, each spatial point \(x_j\) is associated with a positive quadrature weight \( w_j \) that represents the local spacing:
\begin{equation}
w_j =
\begin{cases}
\dfrac{x_{2} - x_{1}}{2}, & j = 1, \\[8pt]
\dfrac{x_{j+1} - x_{j-1}}{2}, & 2 \le j \le N_x - 1, \\[8pt]
\dfrac{x_{N_x} - x_{N_x-1}}{2}, & j = N_x.
\end{cases}
\label{eq:trapezoidal_weights}
\end{equation}
The weights are collected into a diagonal matrix:
\[
\umat{W} = \mathrm{diag}(w_1, w_2, \dots, w_{N_x}).
\]

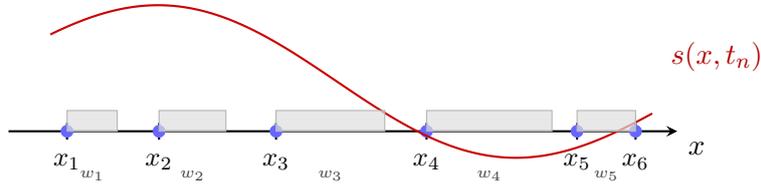
\begin{figure}[h!]
\centering
\begin{tikzpicture}[scale=1.1, >=stealth]

\draw[->, thick] (0,0) -- (8,0) node[below right] {$x$};

\foreach \x/\name in {0.7/{x_1},1.8/{x_2},3.2/{x_3},5/{x_4},6.8/{x_5},7.5/{x_6}}{
  \draw[thick] (\x,0.1) -- (\x,-0.1);
  \node[below] at (\x,-0.15) {\small $\name$};
  \filldraw[blue!60] (\x,0) circle (2pt);
}

\draw[smooth, thick, red!80!black, domain=0.5:7.7, samples=100]
  plot(\x, {0.7*sin(0.8*\x r)+0.3*cos(0.4*\x r)+0.6});

\node[red!70!black, right] at (7.8,0.9) {$s(x,t_n)$};

\foreach \x/\w/\k in {0.7/0.6/1,1.8/0.8/2,3.2/1.3/3,5/1.5/4,6.8/0.7/5}{
  \fill[gray!25,opacity=0.7] (\x,0) rectangle ++(\w,0.25);
  \draw[gray!60] (\x,0) rectangle ++(\w,0.25);
  \node[gray!50!black, below] at ({\x+0.5*\w},-0.35)
       {\tiny $w_{\k}$};
}

\end{tikzpicture}

\caption{Illustration of non-uniform spatial sampling points \(x_j\) and associated quadrature weights \(w_j\) used in the weighted C.O.D. formulation. The shaded rectangles represent the local spatial weights used to approximate the integral.}
\label{fig:nonuniform_grid_cod}
\end{figure}

\subsection*{1. Modal decomposition via C.O.D. on a non-uniform grid}

We still construct the same way the analytic signal (see III.1), as well as the analytic signal matrix:

\begin{equation}
\umat{Z} = \umat{S_c}^T \in \mathbb{C}^{N_x \times N_t}.
\end{equation}

To incorporate the non-uniform spacing, the spatial inner product between two complex spatial vectors
\(\uvec{u}, \uvec{v} \in \mathbb{C}^{N_x}\)
is defined as a weighted discrete quadrature:
\begin{equation}
\langle \uvec{u}, \uvec{v} \rangle_w = \uvec{u}^\dagger \umat{W} \uvec{v}
= \sum_{j=1}^{N_x} w_j\, u_j^\ast v_j.
\end{equation}

The corresponding temporal covariance matrix is:
\begin{equation}
\umat{R_w} = \frac{1}{N_t}\, \umat{Z}\, \umat{W}\, \umat{Z}^\dagger \in \mathbb{C}^{N_x \times N_x}.
\end{equation}
The weighting matrix \( \umat{W} \) ensures that spatial correlations are evaluated consistently with the non-uniform grid. The matrix \( \umat{R_w} \) remains Hermitian and positive semi-definite.

We then solve the generalized eigenvalue problem:
\begin{equation}
\umat{R_w}\, \uvec{\phi}_j = \lambda_j^e\, \uvec{\phi}_j,
\end{equation}
yielding:
\begin{itemize}
    \item eigenvalues \( \lambda_j^e \in \mathbb{R}_+ \) (modal energies),
    \item eigenvectors \( \uvec{\phi}_j \in \mathbb{C}^{N_x\times 1} \) (spatial modes).
\end{itemize}

The temporal coefficients are obtained by projection using the weighted inner product:
\begin{equation}
\uvec{a}_j = \uvec{\phi}_j^\dagger \umat{W} \umat{Z},
\end{equation}
or equivalently,
\begin{equation}
\umat{A} = \umat{\Phi}^\dagger \umat{W} \umat{Z},
\end{equation}
with \( \umat{\Phi} = [\uvec{\phi}_1, \uvec{\phi}_2, \dots] \).

The reconstruction then reads:
\begin{equation}
\umat{S_c} = (\umat{A}^T \umat{\Phi}^T) \approx \sum_{j=1}^{N_x} \uvec{a}_j^T \uvec{\phi}_j^T.
\end{equation}

\subsection*{2. Modal energies and orthogonality}

Each eigenvalue quantifies the weighted energy of the corresponding mode:
\begin{equation}
\lambda_j^e = \frac{1}{N_t} \|\uvec{a}_j\|^2
= \frac{1}{N_t} \sum_{n=1}^{N_t} |a_{j}(t_n)|^2.
\end{equation}

The spatial modes are orthonormal under the weighted inner product:
\begin{equation}
\uvec{\phi}_j^\dagger \umat{W} \uvec{\phi}_k =
\begin{cases}
1, & j = k, \\
0, & j \neq k.
\end{cases}
\end{equation}

This ensures that energy is conserved consistently with the continuous-space formulation:
\[
\int |s_c(x,t)|^2 \, dx
\;\approx\;
\uvec{s_c}^\dagger \umat{W} \uvec{s_c}.
\]

\subsection*{3. Reconstruction of the real-valued field}

To recover the physical real-valued field, we still take the real part:
\begin{equation}
\umat{S} \approx \Re[\,\umat{S_c}\,] = \Re[\,\umat{Z}^T\,].
\end{equation}

\subsection*{4. Travelling index on non-uniform grids}

The computation of the travelling index for each mode remains unchanged, except that all spatial inner products involved in the Gram matrix are now weighted:
\begin{equation}
\underline{\underline{G}}_j =
\begin{bmatrix}
(\underline{\phi}_j^{\Re})^T \umat{W} \underline{\phi}_j^{\Re} &
(\underline{\phi}_j^{\Re})^T \umat{W} \underline{\phi}_j^{\Im} \\[4pt]
(\underline{\phi}_j^{\Im})^T \umat{W} \underline{\phi}_j^{\Re} &
(\underline{\phi}_j^{\Im})^T \umat{W} \underline{\phi}_j^{\Im}
\end{bmatrix}.
\end{equation}
The travelling index is still given by:
\begin{equation}
\alpha_j  =
\sqrt{\frac{a + b - \sqrt{(a - b)^2 + 4c^2}}{a + b + \sqrt{(a - b)^2 + 4c^2}}},
\end{equation}
with \(a,b,c\) computed from the weighted Gram matrix entries.

\bigskip
In summary, the non-uniform grid formulation preserves all properties of the standard discrete C.O.D., provided that the spatial weighting matrix \( \umat{W} \) is systematically applied to every spatial inner product and correlation.

\subsection*{5. Example}

\begin{figure}[ht!]
    \centering
    \includegraphics[scale=1]{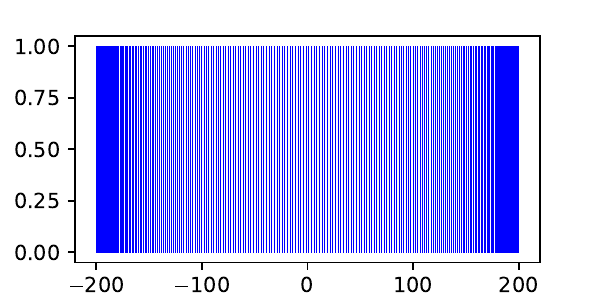}
    \caption[]{Representation of the spacing of the points on the non-uniform grid}
    \label{chebych}
\end{figure}

The full calculus are chosen to be left aside. The following figures are just here to illustrate the Complex Orthogonal Decomposition on non-uniform grids. These examples come from \textit{test\_COD\_ex\_1\_nu.py}, that is taking exactly the first example of the sloshing wave but with a sinusoidal grid (Tchebychev reparted nodes) as represented on Figure \ref{chebych}. We keep the same parameters, only the grid is modified.

\begin{figure}[ht!]
                \begin{minipage}[c]{.46\linewidth}
     \centering
    \includegraphics[scale=0.55]{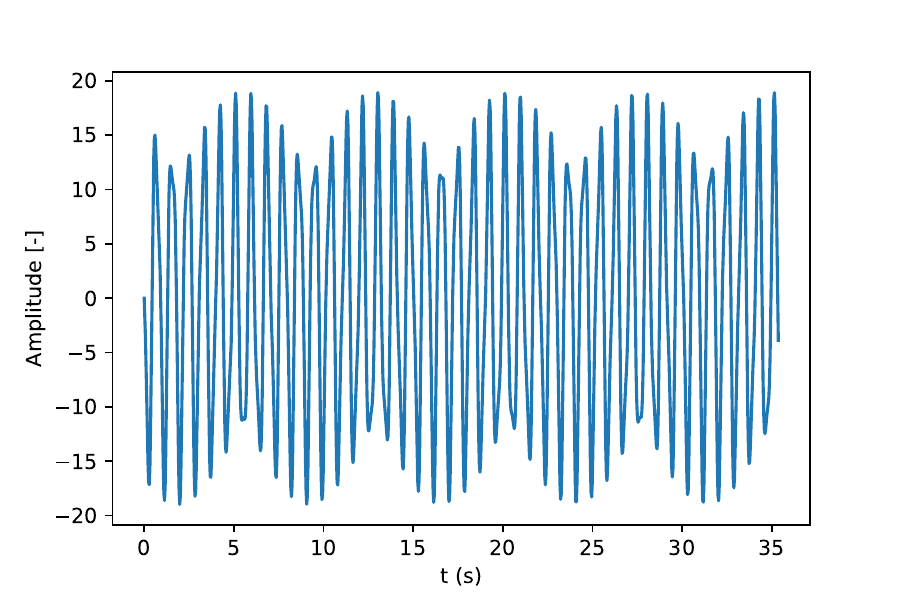}
                \end{minipage}
                \hfill%
                \begin{minipage}[c]{.46\linewidth}
                    \centering
                    \vspace{-0.1mm}  
                    \includegraphics[scale =0.55]{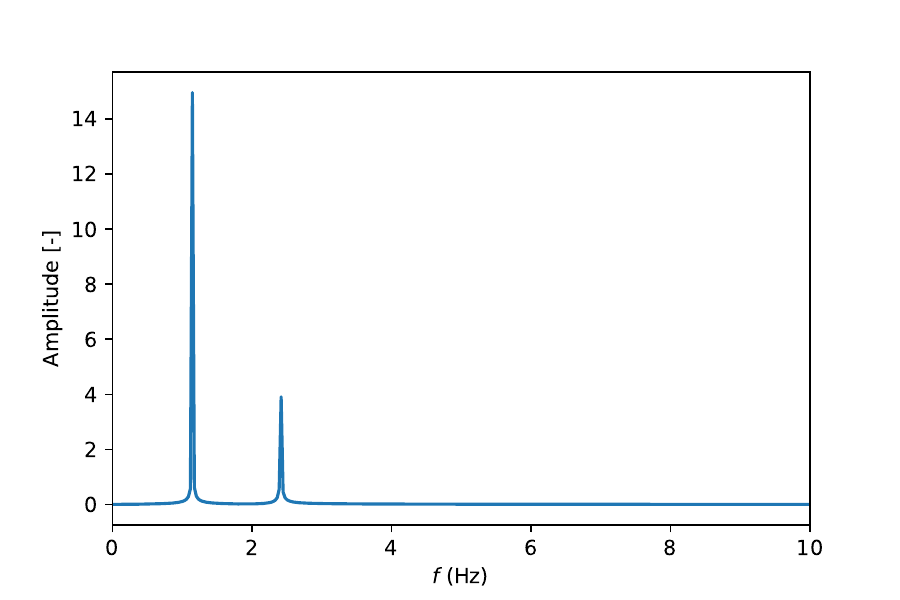}

                \end{minipage}
            \caption[]{Time signal (\textit{left}) and Fourier spectrum (\textit{right}) at $x=-200$ mm (on the wall of the tank, antinode of oscillations for both modes)}
                \label{Fig3bis}
            \end{figure}

As a recall, Figure \ref{Fig3bis} shows the signal at one point and also the associated time Fourier transform of this signal. There is no difference with the example of section \ref{IV}, since it is only a point information.

\begin{figure}[ht!]
                \begin{minipage}[c]{.46\linewidth}
     \centering
    \includegraphics[scale=0.55]{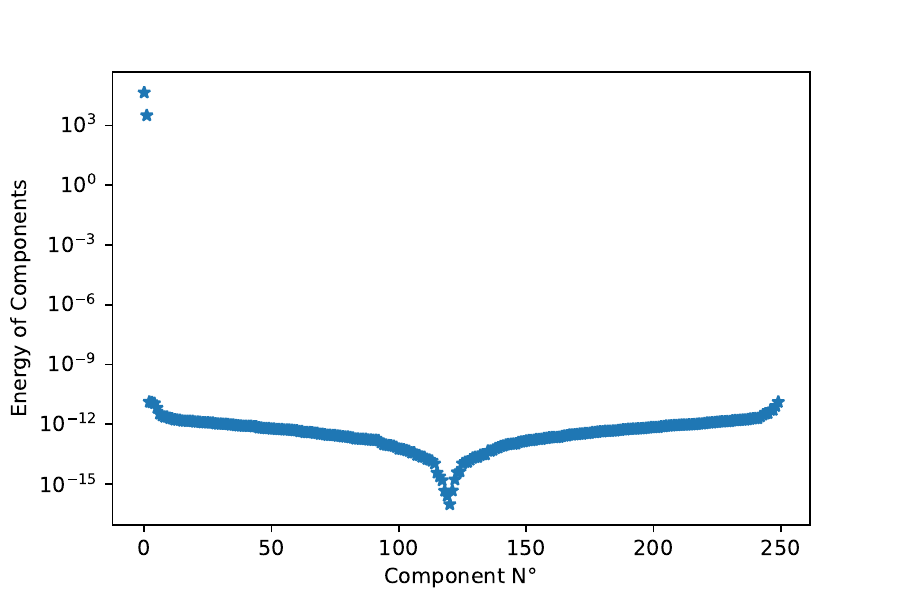}
                \end{minipage}
                \hfill%
                \begin{minipage}[c]{.46\linewidth}
                    \centering
                    \vspace{-0.1mm}  
                    \includegraphics[scale =0.55]{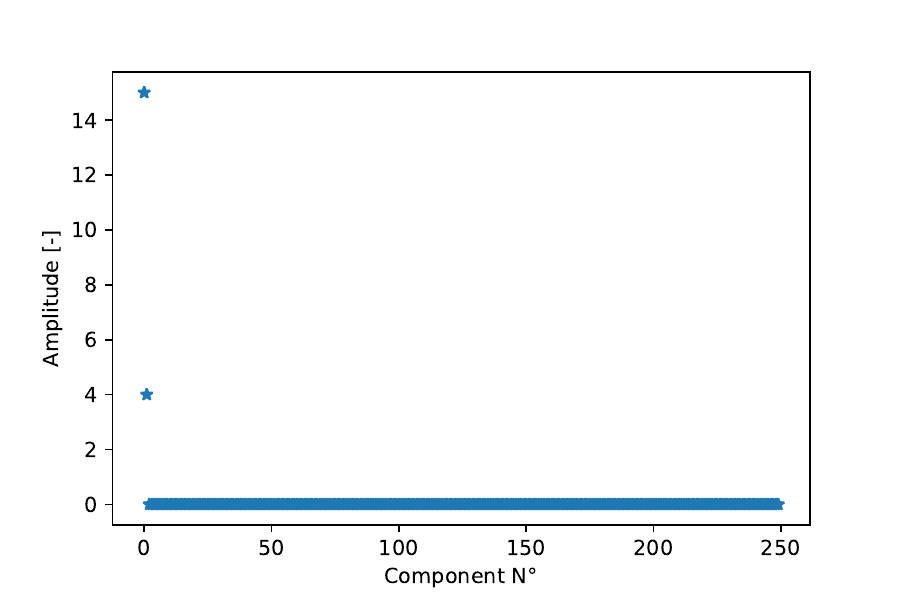}

                \end{minipage}
            \caption[]{Energy of the components $\lambda^{e}_{i}$ (\textit{left}) and corresponding oscillations amplitudes $A_i$ (\textit{right}) in the case of non-uniform spatial grid}
            \label{fig4bis}
            \end{figure}

On Figure \ref{fig4bis}, we display the results of the weighted C.O.D. with the energies and amplitudes of the components of the decomposition. No significant difference is seen, demonstrating in this case the effectiveness of the weighted decomposition in the case of a non-uniform space grid. 

\begin{figure}[ht!]
                \begin{minipage}[c]{.46\linewidth}
     \centering
    \includegraphics[scale=0.55]{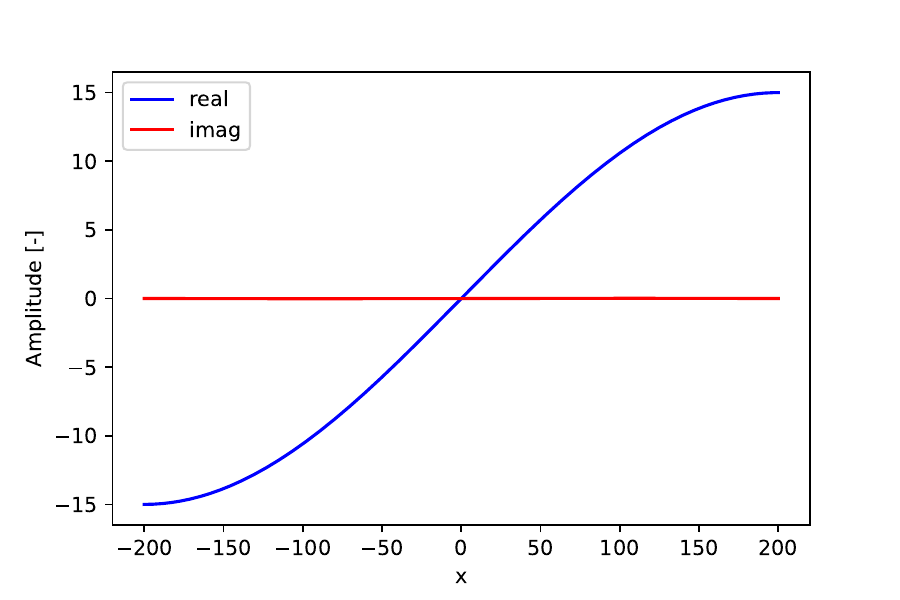}
                \end{minipage}
                \hfill%
                \begin{minipage}[c]{.46\linewidth}
                    \centering
                    \includegraphics[scale =0.55]{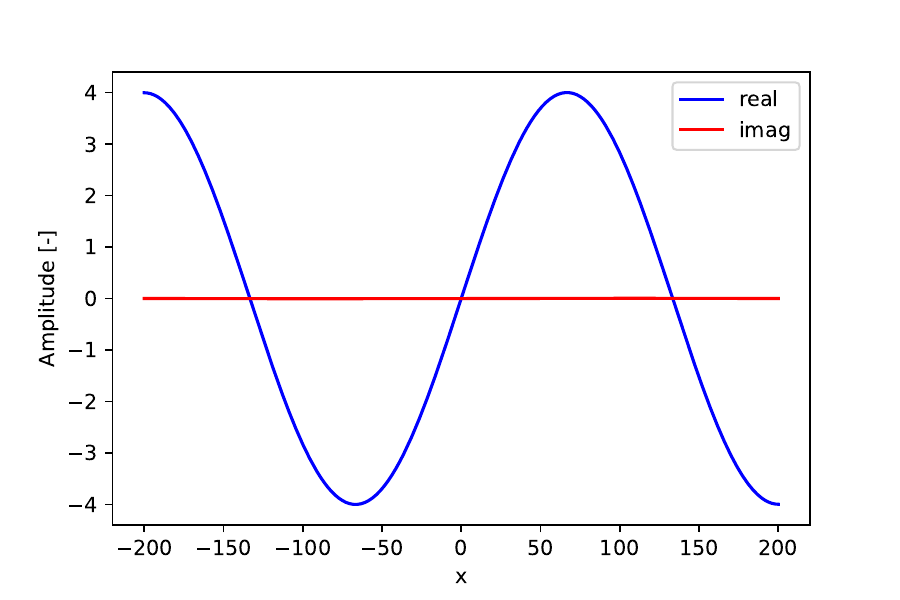}
                \end{minipage}
            \caption[] {$\phi_i^{\Re}(x)$ and $\phi^{\Im}_i(x)$ for $i=1$ (\textit{left}) and $i=2$ (\textit{right}) in the case of a non-uniform grid}
            \label{fig5bis}
            \end{figure}

\begin{figure}[ht!]
    \centering
    \includegraphics[scale=0.65]{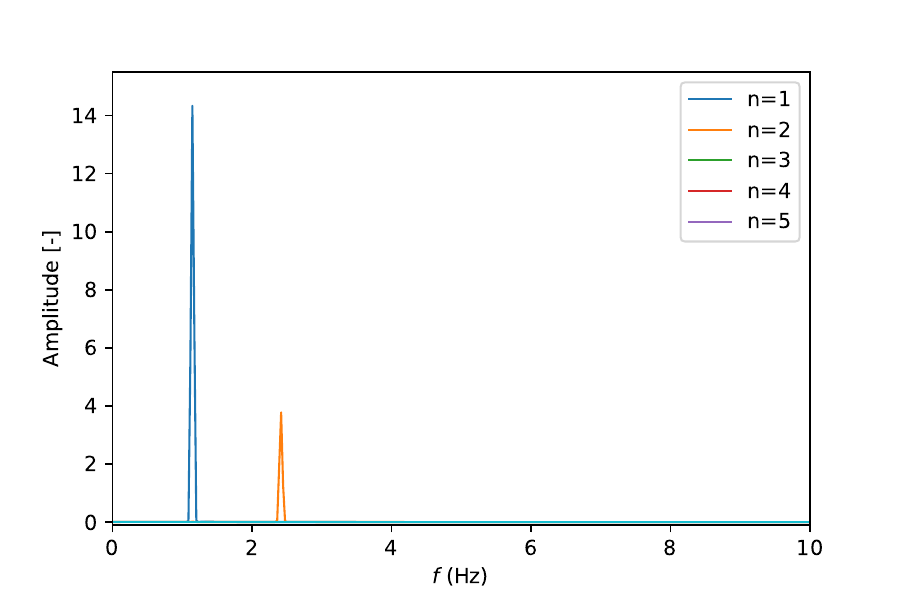}
    \caption[]{Fourier spectra of the 5 first components of the C.O.D. (non-uniform grid)}
    \label{fig6bis}
\end{figure}

Figure \ref{fig5bis} shows the spatial form of each of the two non-zero components : it recovers the expected spatial forms. The same conclusion can be drawn speaking about the temporal behaviour when looking at Figure \ref{fig6bis}. Indeed, time Fourier spectra show that both frequency are recovered and separated.

\newpage

\newpage

\newpage

\bibliography{COD} 
\bibliographystyle{ieeetr}

\addcontentsline{toc}{part}{References} 

\end{document}